\documentclass[12pt]{article}

\usepackage[dvipsnames]{xcolor}
\usepackage{amsmath}
\usepackage{amssymb}
\usepackage{dsfont}
\usepackage[english]{babel}
\usepackage[utf8]{inputenc}
\usepackage{times}
\usepackage{graphics}
\usepackage{graphicx}
\usepackage[T1]{fontenc}
\usepackage[official,right]{eurosym}
\usepackage{url}
\usepackage{eso-pic}

\usepackage{authblk}
\usepackage[colorlinks=true, allcolors=blue]{hyperref}
\usepackage{array} 
\usepackage{caption}
\usepackage{booktabs}
\usepackage{multirow}
\usepackage{enumitem}
\usepackage{xcolor}
\usepackage[most]{tcolorbox}
\usepackage{geometry}
\usepackage{tabularx}
\usepackage{pdflscape}
\usepackage{pgfplots}
\pgfplotsset{compat=1.17}
\usepackage{pgfplotstable}
\usepackage{tikz}
\usetikzlibrary{shapes.multipart, arrows.meta, positioning}
\usepackage{qtree}

\usepackage{tikz}
\usetikzlibrary{mindmap}
\usetikzlibrary{shapes.geometric, arrows}

\tikzstyle{startstop} = [rectangle, rounded corners, minimum width=3cm, minimum height=1cm,text centered, draw=black, fill=gray!20]
\tikzstyle{process} = [rectangle, minimum width=3cm, minimum height=1cm, text centered, draw=black, fill=blue!10]
\tikzstyle{decision} = [diamond, minimum width=3cm, minimum height=1cm, text centered, draw=black, fill=red!10]
\tikzstyle{arrow} = [thick,->,>=stealth]

\definecolor{clearblue}{rgb}{0.3, 0.6, 1.0}

\title{Explainability and justification of automatic-decision making -- A conceptual framework and a practical application}

\author[1]{Sarra Tajouri}
\author[2]{Yves Meinard}
\author[1]{Alexis Tsoukiàs}
\author[3]{Thierry Kirat}

\affil[1]{LAMSADE-CNRS, Université Paris Dauphine - PSL}
\affil[2]{Aix Marseille Univ, CNRS, CGGG, Centre Gilles Gaston Granger, Aix-en-Provence, France}
\affil[3]{TRIANGLE-CNRS, ENS Lyon, Université Lumière-Lyon 2, Sciences Po Lyon, Université Jean Monnet-Saint-Etienne}

\date{February, 2026}

\begin{document}

\thispagestyle{empty}

\enlargethispage*{8cm}
 \vspace*{-38mm}

\AddToShipoutPictureBG*{\includegraphics[width=\paperwidth,height=\paperheight]{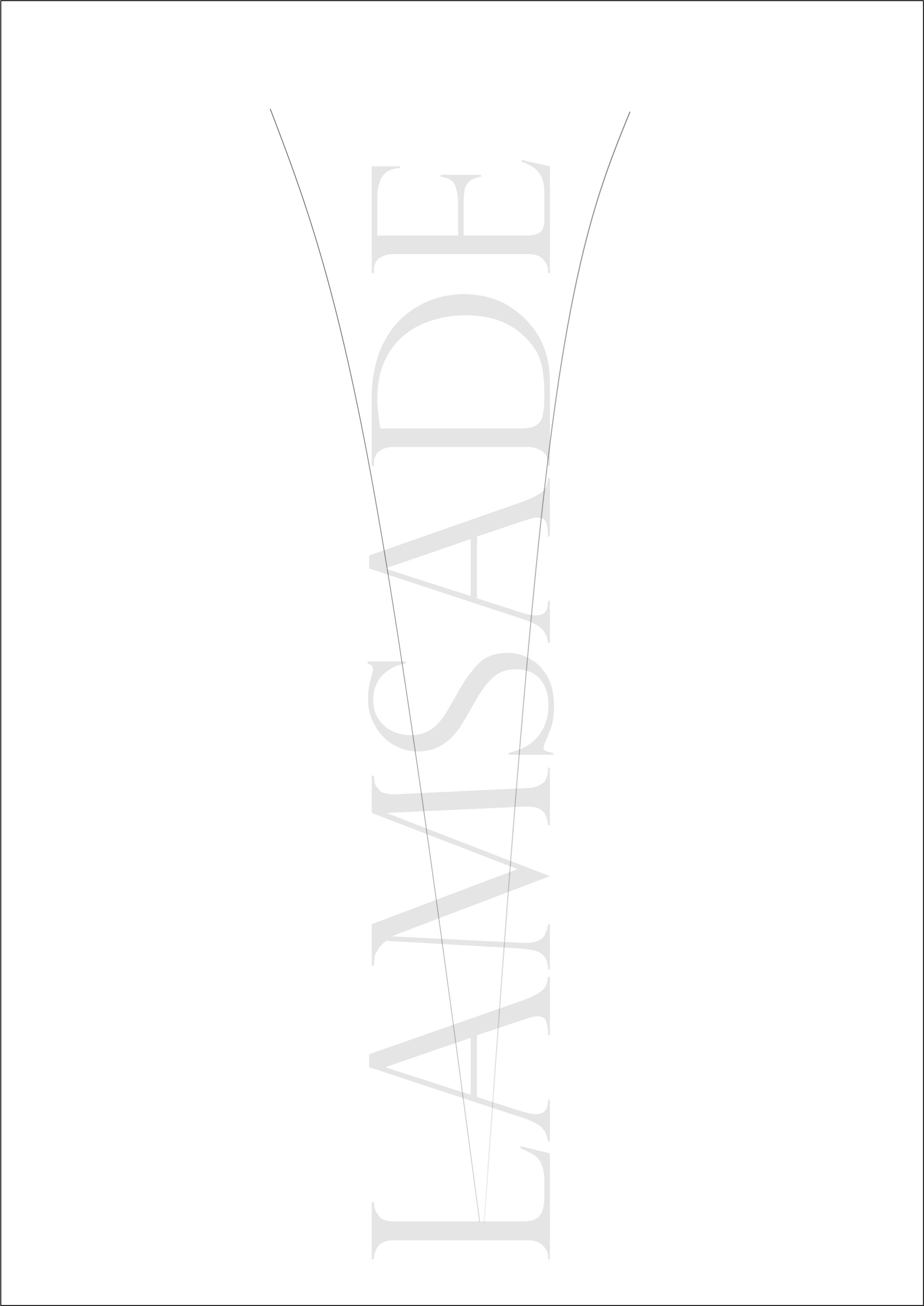}}

\begin{minipage}{24cm}
 \hspace*{-28mm}
\begin{picture}(500,700)\thicklines
 \put(60,670){\makebox(0,0){\scalebox{0.7}{\includegraphics{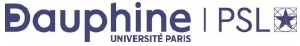}}}}
 \put(60,70){\makebox(0,0){\scalebox{0.07}{\includegraphics{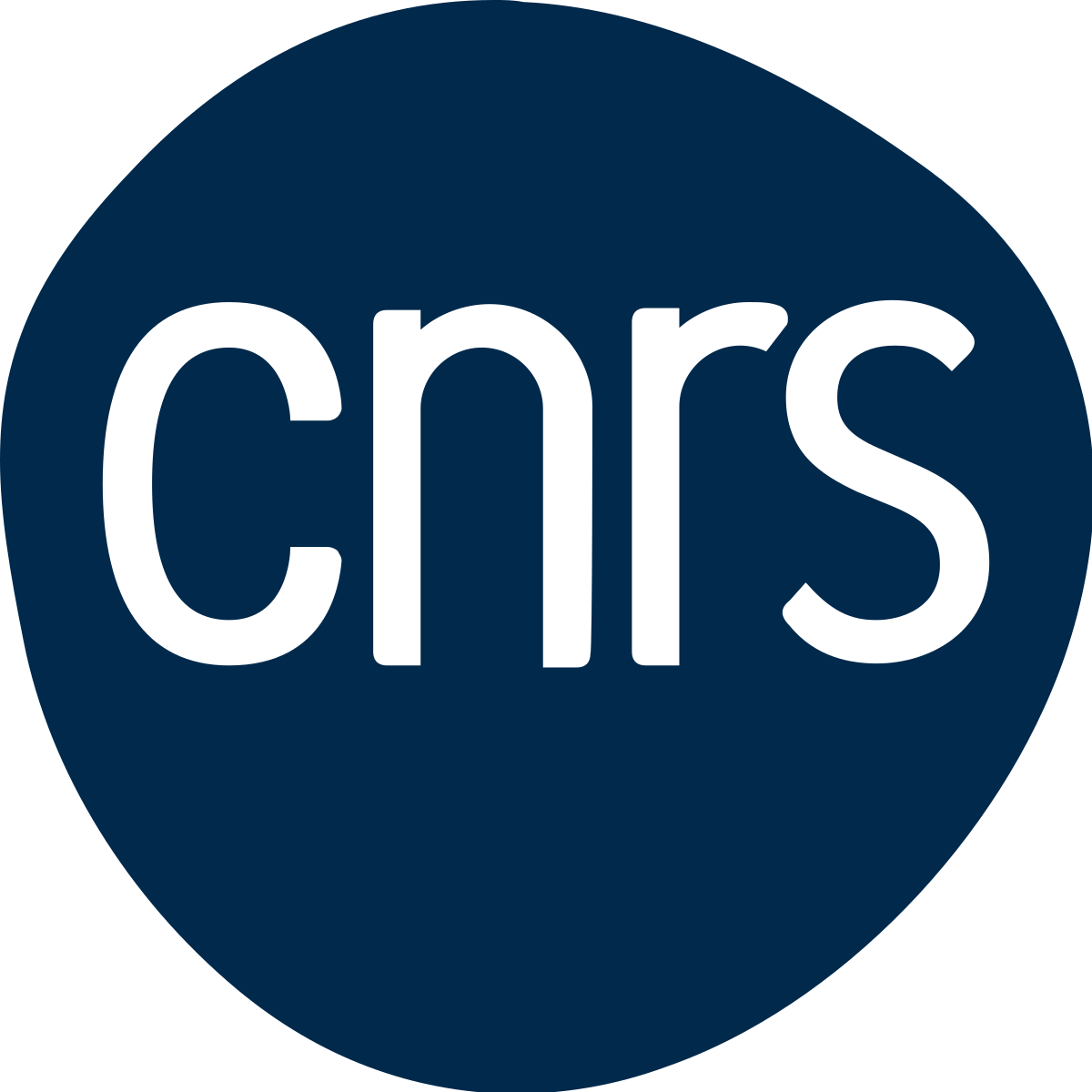}}}}
 \put(320,350){\makebox(0,0){\Huge{CAHIER DU \textcolor{BurntOrange}{LAMSADE}}}}
 \put(140,10){\textcolor{BurntOrange}{\line(0,1){680}}}
 \put(190,330){\line(1,0){263}}
 \put(320,310){\makebox(0,0){\Huge{\emph{413}}}}
 \put(320,290){\makebox(0,0){February, 2026}}
 \put(320,210){\makebox(0,0){\Large{Explainability and justification of }}}
 \put(320,190){\makebox(0,0){\Large{automatic-decision making}}}
 \put(320,170){\makebox(0,0){\Large{A conceptual framework and a practical application}}}
 \put(320,100){\makebox(0,0){\Large{Sarra Tajouri, Yves Meinard}}}
  \put(320,80){\makebox(0,0){\Large{ Alexis Tsoukiàs, Thierry Kirat}}}

 \put(320,670){\makebox(0,0){\Large{\emph{Laboratoire d'Analyse et Mod\'elisation}}}}
 \put(320,650){\makebox(0,0){\Large{\emph{de Syst\`emes pour l'Aide \`a la D\'ecision}}}}
 \put(320,630){\makebox(0,0){\Large{\emph{UMR 7243}}}}
\end{picture}
\end{minipage}

\newpage

\addtocounter{page}{-1}

\maketitle

\abstract{Explainability of algorithmic decision-making systems is both a regulatory objective and an area of intense research. The article argues that a crucial condition for the acceptability of algorithmic decision-making systems is that decisions must be justified in the eyes of their recipients. We make a clear distinction between explanation and justification. Explanations describe how a decision was made, while justifications give reasons that aim to make the decision acceptable. We propose a conceptual framework of explanations and justifications, based on Habermas’s theory of communicative action and Perelman's New Rhetoric theory of law. This framework helps to analyze how different forms of explanation can support or fail to support justification. We illustrate our approach with a case study on university admissions in France. }

\newpage

\section{Introduction}
The use of decision-support systems has become increasingly prevalent in contemporary societies. Many people are subject to decisions made by these systems on a regular basis, often without even realizing it \cite{danaher2016threat, waldman2019power}. Their influence spans a broad spectrum of contexts, from seemingly trivial everyday interactions, such as personalized recommendations for movies, food, or consumer products, to high-stakes decisions involving access to social benefits, creditworthiness assessment, healthcare diagnosis, and even predictive policing \cite{eubanks2018automating, haque2024are, saxena2022unpacking, stevenson2018assessing}.

In some particularly high-stakes domains, such as access to public or private resources, these decision-support systems can have profound impacts, including unintended consequences on people's lives and even communities \cite{liu2018delayed}. For instance, credit scoring algorithms used by banks to assess loan applications can (presumably inadvertently) discriminate against applicants from minorities or with low-income backgrounds, denying them access to credit and perpetuating cycles of poverty \cite{bono2021algorithmic, hurlin2022fairness}. Another telling example is the one of predictive policing algorithms that identify ``high-crime'' areas and thereby can lead to over-policing of marginalized neighborhoods, exacerbating systemic racism \cite{bates2024technology}. Similarly, in healthcare, algorithms can assist in diagnosing diseases or prioritizing patients for treatment, directly affecting patient care outcomes \cite{agarwal2023addressing, parikh2019addressing, obermeyer2019dissecting}. Recommendation systems on social media platforms also shape public discourse by influencing the information that people can access \cite{cinus2022effect, lanzetti2023impact}.

In most cases, such systems or tools are called ``decision tools'' through misuse of language, since what they really do is produce a recommendation that an agent or agents can use or ignore when making their decision. However, the fact that most of these systems do not make decisions themselves does not make those decisions any less consequential.

Algorithmic systems hence actively shape society. They participate in social and economic structures, entrench norms (e.g., norms of creditworthiness or risk assessment; \cite{fourcade2013classification, kasy2021fairness}), and even mediate interactions \cite{gillespie2018custodians}, thereby displaying attributes that typically rather characterize social actors, such as agency and influence on social relationships. 

Consequently, it is now largely accepted that these systems must align with certain norms and expectations, typically expressed using terms like ``fairness'' or ``explainability'', and they are increasingly subject to accountability requirements similar to those expected from human social actors \cite{ananny2018seeing}. However, algorithms do not function in the same way as human decision-makers and, crucially, people perceive and judge algorithmic and human decisions in distinct ways \cite{green2019principles,mok2023people}. Among key-differences, a stark contrast, which is unusual in human decision contexts, often characterizes automated decisions. While the accuracy and efficiency of human decisions often go hand in hand with decision-makers' ability to account for the way decisions were made, automated systems often manage to perform highly accurately and efficiently some tasks without anyone being able to fully understand the details of how these tasks were performed. This difficulty is compounded when such systems produce recommendations that have multifarious, diffuse impacts, such as reinforcing existing societal inequalities \cite{caliskan2017semantics}, or when they reflect the biases and assumptions of the model developers \cite{mitchell2018prediction}. The challenge of adapting to automated decisions requirements originally applicable to human decisions has prompted legislators to elaborate legal frameworks and regulations intended to guide the use of at least partly automated decision systems \cite{gdpr, aiact}.

A related challenge echoes the above clarification that such systems are typically not decision makers, but recommendation systems. Because what they produce is recommendations, the usability of such systems and, in fine, their effectiveness, depend on recommendations being followed by decision-makers and their consequences accepted by people they affect. The challenge, which is also typically addressed in discussions articulated in the terms just mentioned (``explanability'', ``fairness'', etc.) is hence to foster the appropriation and acceptability of such systems.

In this article, we elaborate a framework designed to streamline and rationalize such attempts at providing a normative structure to the use of, at least partly, automated decision support systems. Most of the attempts in that direction in the existing literature revolve around the terms ``explanation'' and ``justification'' (or associated terms such as ``explanability'' and the like), but these terms are given different meanings in different contexts, resulting in a blurry and unclear picture. To address this predicament, we propose a clarification of the corresponding notions.

The remainder of this paper is organized as follows. Section \ref{section2} reviews the academic literature to map the meanings given to the terms ``explanation'', ``justification'' and the like, particularly in their application to, at least partly, automated decision processes. As a preliminary step in this clarification, we briefly examine how European legislation addresses these concepts, particularly the prominence of explanation and the relative absence of any formal justification requirement. This legal backdrop helps highlight both the normative expectations currently placed on algorithmic systems and the conceptual gaps that remain unaddressed in regulation, thus reinforcing the need for a more precise theoretical account. We then propose in section \ref{section3} our own conceptual framework to clarify the concepts of explanation and justification, based on the work of contemporary philosopher Jurgen Habermas \cite{habermas1987com} and legal theorist Chaïm Perelman \cite{perelman1958traite}. The last section \ref{section4} proposes an application of this conceptual framework in a concrete, real-life case-study.

\section{``Explanation'' and ``justification'' in the existing legislation and literature}
\label{section2}
As mentioned above, this section examines the main features of European legislation on explainability and highlights a gap with regard to the issue of justification. It then goes on to discuss in greater detail the existing types of explanation and their compatibility with justification. 


\subsection{Explainability requirements and justification gaps in European legislation}
A prominent legal framework in the domain is the GDPR (General Data Protection Regulation) \cite{gdpr}, which was enacted by the European Union in 2018. The GDPR includes a ``right to explanation'', but fails to explicitly define what it calls an ``explanation''. It establishes ``transparency" requirements for automated decision-making. Articles 13 and 14 mandate that individuals should receive 
``meaningful information about the logic, significance, and consequences of decisions" based on automated systems. Article 22 grants individuals the ``right not to be subject to fully automated decisions", while Recital 71 emphasizes the need for ``fair" and ``transparent" processing, suggesting a right to understand and challenge algorithmic decisions. However, some researchers \cite{edwards2018slave,wachter2017right} question the feasibility of such a right, criticizing its lack of precise language. Additionally, it is important to note that this regulation applies only to fully automated decision-making systems, which are relatively rare in practice. Indeed a distinction must be made between a ``decision" and a ``recommendation". The former refers to an allocation of resources (of any type) to one or more tasks, formally defined as a partition of a set (of ``possible allocations'') which actually changes the state of the system upon which the allocation acts \cite{ColorniTsoukias2024}. By contrast, the latter is a mere suggestion that a decision (in the above sens) should be made. 

More recently, the AI Act \cite{aiact}, proposed by the European Commission in 2021 and enacted in June, 2024, targets high-risk AI systems, demanding transparency and clear documentation of their decision-making processes. This regulation aims to ensure that both developers and users of AI systems offer comprehensible ``explanations'', enhancing accountability and safeguarding individual rights in critical sectors such as healthcare, finance, and public administration. It mandates that providers implement a risk management system understood as a \emph{continuous and iterative process to identify, analyze risks} that may arise from both intended use and unintended consequences (Article 9). 

In order to maintain human control, the AI Act enforces human oversight mechanisms (Article 14) and makes mandatory for AI providers to both continuously monitor system performance and report incidents (Article 61). Finally, strict enforcement measures, including financial fines (Articles 85-92), ensure that AI developers remain legally accountable for the societal impact of their systems. Even though this regulation is one of the most significant of its kind concerning high-stakes AI systems, it does not provide any definition of ``explanation'' at any point in the text.

Whereas the GDPR contains no explicit requirement for justification, the AI Act uses the language of ``accountability'', a legal term closely related to justification. In legal theory, to justify a decision often entails showing that it abides by established standards — precisely the kind of reasoning that underlies accountability in law \cite{Goltzberg2013argumentation}. Yet, despite this conceptual proximity, the AI Act notably refrains from using the term ``justification'' itself or from framing obligations in those terms.

This raises the question: why does the legislation stop short of requiring justification, despite acknowledging the importance of accountability? 
The absence of explicit justification requirements suggests a regulatory gap, or at least an ambiguity, in how algorithmic decisions are expected to meet standards of legitimacy.

To address this ambiguity, we turn to the propositions of the academic literature to clarify how the term ``explainability'' and related notions are used. As a brief historical note, XAI (acronym for eXplainable Artificial Intelligence) was introduced by the Defence Advanced Research Projects Agency (DARPA) in 2016, which called for innovative research proposals in AI around techniques producing ``interpretable'' systems that could be understood and trusted by human beings \cite{gunning2016explainable}. The term ``XAI'' has since been widely used to refer to models that provide ``insights'' into how they arrive at their outcomes \cite{arrieta2020explainable}. The terms ``interpretability'' and ``explainability'' are also widely used in the literature, sometimes interchangeably, to describe the goal of making systems’ decisions clearer to human users. By contrast, some authors argue that the two terms should be distinguished, using the term \emph{interpretability} to refer to the availability of a human-understandable description of the system's internals, while anchoring ``explainability'' in the notion of ``explanation'', understood as a decision maker/user interface providing the later with details about the model functioning \cite{doshi2017towards, gilpin2018explaining}.

In a review of opportunities and risks related to the use of algorithmic decision systems (ADS) commissioned by the European Parliament,  Castellucia et al.~\cite{castelluccia2019understanding} make a valuable effort to clarify such terminologies. ADS are defined as computational systems that assist (or replace) human decision-making by processing large amounts of data to infer correlations. This report is particularly relevant to our study because it articulates a structured set of desiderata. The definition they advance can be summarized as follows:
\begin{itemize}[left=0pt, topsep=0pt]
    
    \item ``understandability'' is the possibility to provide understandable information (for a human being) about the link between the input and the output of the ADS. 
    
    \item ``transparency'' is a form of understandability that implies access to information about the internal workings of the algorithm (code, documentation, training data-sets...). It stands in contrast to ``opacity'', which characterizes black-box models -- systems whose internal workings are either too complex to interpret due to a high number of parameters or deliberately hidden. 
    
    \item ``explainability'' is a form of understandability that is defined as the availability of explanations about the ADS. In contrast to transparency, explainability requires the delivery of meaningful information beyond the ADS's descriptive artifacts.
     \vspace{0.25cm}
     
    Therefore, understandability refers to the potential to provide comprehensible information about the system, whether in the form of raw information (e.g., code, documentation) as in transparency, or through the availability of meaningful explanations as in explainability.
    
    \item ``explanations'' are defined technically, either by their form (decision trees, histograms...) or by their type: either operational (informing about how the system actually works), logical (informing about the logical relationships between inputs and results) or causal (informing about the causes for the results), global (about the whole algorithm) or local (about specific results). Thus the availability of any kind of explanations makes the ADS ``explainable''.

    \item ``interpretability'': Castelluccia et al. \cite{castelluccia2019understanding} do not offer any consistently formalized definition of interpretability, as they consider it to be very closely related to explainability. They adopt Lipton’s \cite{lipton2018mythos} formulation, which states that “explanation is post-hoc interpretability”. In general, a system is described as interpretable either when it relies on a simple, transparent algorithm that can be readily understood, or when it produces explanations that render its decisions comprehensible. In that sense, ``interpretability'' is a synonym of understandability.
    
    \item ``accountability" refers to the obligation for an agent or system to specify justifications for its decisions, along with the possibility of facing sanctions if those justifications are deemed inadequate. This definition is based on the work of Binns \cite{binns2018algorithmic} that frames accountability through the democratic political ideal \emph{public reason} as defined by Rawls and Habermas \cite{rawls1997idea, habermas1993justification}. According to this view, justifications should be based on epistemic and normative standards that can be reasonably accepted by all members of a democratic society.
\end{itemize}   

%

In the above definitional framework, explanations, of different types, hence serve primarily to enhance understandability. Justification, by contrast, becomes relevant specifically in the context of accountability. 
In what follows, we expose some typologies of explanations and justifications that can be found in the literature.
  
\subsection{Existing typologies of explanation in the literature}
\label{extyp}
There exist numerous types of explanation methods extensively discussed in the literature, ranging from interpretability techniques for white-box models to post-hoc approximations for black-box models (see \cite{arrieta2020explainable,burkart2021survey,guidotti2018survey,gunning2019darpa,nauta2023anecdotal}). For space reasons,  we leave aside the vast philosophical literature on the classical explanation/understanding dichotomy \cite{vonwright1971,ricoeur2014memoire}, because it refers to a historically situated debate in the epistemology of historical disciplines, which falls beyond our scope ; although its topic is to some degree akin to our own subject matter, the even vaster philosophical literature on the reason/cause dichotomy \cite{Lowe2009} is left aside for the same reason.

Hence, we do not claim to survey the literature exhaustively. We rather illustrate the main typologies found in the literature, each defined by different aspects such as what they cover, how they are structured, who they are meant for, and what kind of thinking they use. We introduce a running example to illustrate key features of the various typologies and key-aspects on which they differ.

\begin{tcolorbox}[colback=green!5!white, colframe=green!50!black, title=Example of ADS: a credit scoring system]

Let us consider a credit risk assessment system that estimates the probability of default for each loan application. This probability is computed by a predictive model trained on historical data of past borrowers. The model takes as input features such as income level, employment status, existing debt, and credit history, and outputs a value between 0 and 1 representing the estimated likelihood that the applicant will fail to repay the loan (therefore 0 is preferred to 1). Typically, the decision-maker (DM) uses this probability to make a binary decision: if it exceeds a predefined threshold (let's note it $t$), the application is rejected; otherwise, it is approved.
\end{tcolorbox}

\subsubsection{Types of explanations sorted by scope}

The scope of an explanation (or \emph{scale} \cite{mohseni2021multidisciplinary}) is the level at which the explanation is provided.
The most widespread categorization distinguishes \emph{local} vs. \emph{global} explanations. Local explanations focus on individual recommendations, while global explanations account for the entire system behavior.  

\begin{tcolorbox}[colback=green!5!white, colframe=green!5!white, breakable, enhanced]

\textbf{A local explanation} targets a \textbf{single prediction} $f(x)$. Usually, we aim to identify the \emph{main contributing features} that led to the model’s output for a specific instance $x$. For instance, feature attribution methods such as SHAP (SHapley Additive exPlanations) \cite{lundberg2017unified} decompose the prediction:

$f(x) = \phi_0 + \sum_{i=1}^{n} \phi_i$ where

\begin{itemize}
    \item $\phi_0$ is the model’s base value (the expected output across the dataset),
    \item $\phi_i$ is the SHAP value representing the contribution of feature $i$ to the prediction $f(x)$.
\end{itemize}

Let us consider an applicant described by:
$$x = \{\text{income = 900\,}, \ \text{job} = \text{stable}, \ \text{recent\_defaults} = 2\}$$
The model outputs $f(x) = s$ such as $s > t$.

SHAP values for the top features:
\[
\begin{aligned}
\phi_{\text{income}} &= 0.2 \\
\phi_{\text{recent\_defaults}} &= 0.3 \\
\phi_{\text{job}} &= -0.1 \\
\phi_0 &= 0.4 \quad \text{(base approval score)}
\end{aligned}
\]

Final score:
\[
f(x) = 0.4 + 0.2 + 0.3 - 0.1 = 0.8 \Rightarrow \text{denied}
\]

This explanation shows that the decision was mainly driven by \emph{recent defaults} and \emph{low income}, which had stronger negative influence than the positive effect of having a stable job. 

\medskip
\textbf{A global explanation} targets the \textbf{entire model} and aims to provide insight into its overall behavior and logic across all inputs. For instance, one can measure how much each feature generally influences the model’s predictions, a measure commonly referred to as \emph{feature importance}. For example, this can be estimated by aggregating the absolute SHAP values across all data samples.

Let:
\begin{itemize}
    \item \( \phi_j^{(i)} \) be the SHAP value of feature \( j \) for instance \( i \),
    \item \( n \) be the number of data points in the validation set.
\end{itemize}

The global importance of feature \( j \) is computed as:

\[
\text{Importance}(j) = \frac{1}{n} \sum_{i=1}^{n} \left| \phi_j^{(i)} \right|
\]

A higher average absolute value indicates that feature \( j \) consistently contributes more (positively or negatively) to the model's predictions across the dataset.

This method enables ranking features by their overall contribution to the model’s behavior and is often visualized as a bar plot as below:

\begin{tikzpicture}
\begin{axis}[
    xbar,
    bar width=15pt,
    enlargelimits=0.15,
    xlabel={Mean $|\phi|$ (SHAP value)},
    symbolic y coords={recent\_defaults, income, job},
    ytick=data,
    nodes near coords,
    nodes near coords align={horizontal},
    width=105mm,
    height=5cm,
    xmajorgrids=true,
    grid style=dashed
]
\addplot+[fill=blue!50] coordinates {
    (0.55,recent\_defaults)
    (0.35,income)
    (0.22,job)
};
\end{axis}
\end{tikzpicture}

\end{tcolorbox}

\subsubsection{Types of explanations sorted by temporality}

Explanations can be categorized according to when in the decision-making process they are generated: \emph{by-design} or \emph{post-hoc}. \emph{By-design} explanations (also called ``intrinsic'' or ``transparent explanations'') are inherent to the system's architecture and available during development. These explanations arise naturally from models with interpretable structures, such as some decision trees or linear regressions. In contrast, \emph{post-hoc} explanations are generated after model training to interpret already-made decisions. They attempt to explain black-box models after the fact, often by approximating complex models with simpler, more interpretable ones or by analyzing input-output relationships. This temporal distinction has significant implications for explanation fidelity, as by-design explanations typically offer more accurate representations of the actual decision process. Some of the works that discussed this dimension include \cite{adadi2018peeking, guidotti2018survey,  lipton2018mythos, rudin2019stop}.

\begin{tcolorbox}[colback=green!5!white, colframe=green!5!white, breakable, enhanced]
\textbf{Post-hoc explanation:}  
Suppose the original credit scoring system is a black-box model (e.g., a neural network). Due to its complexity and the large number of parameters it relies on, such a model is not inherently interpretable. To explain a specific decision made by this system, we can use a \emph{local surrogate model} — a simpler model trained to approximate the black-box model's behavior in the vicinity of the instance of interest.  

This approach is known as \textbf{LIME} (Local Interpretable Model-agnostic Explanations) \cite{ribeiro2016should}. It works by generating slight perturbations of the original input instance and querying the black-box model for predictions on these perturbed samples. A simple, interpretable model (typically a sparse linear model) is then fitted to these samples.

For example, consider an applicant described by:
\[
x = \{\text{income} = 1000,\ \text{age} = 30,\ \text{recent\_defaults} = 2,\ \text{credit\_history} = \text{poor}\}
\]
The black-box model outputs: $f(x) = s$ such that $s>t$.

LIME fits the following local linear model:
\[
\text{Local Score}(x) = 0.7 \cdot \text{recent\_defaults} - 0.3 \cdot \text{income} + 0.2 \cdot \text{age}
\]

This suggests that the loan was denied primarily due to the applicant's history of defaults. However, it is important to note that this explanation reflects the behavior of the surrogate model in a local neighborhood around \( x \), and not necessarily the global logic of the original black-box model.

\medskip
\textbf{Explanation by design:}  
The model is built to be interpretable from the start. For example, consider a sparse linear model:

\[
\text{Score}(x) = - 0.2 \cdot \text{income} - 0.3 \cdot \text{credit\_history} + 0.5 \cdot \text{recent\_defaults}
\]

Each coefficient reflects the contribution of a feature. A higher income or good credit history decreases the score, while more defaults increase it. If the score exceed the threshold, the loan is denied. This model is globally interpretable without the need for external tools.

\end{tcolorbox}

\subsubsection{By target audience}

Explanations can be distinguished by their intended audience. We may refer to \emph{popularized} explanations when they are intended for lay recipients, and to \emph{expert} explanations when they target technical or professional stakeholders. 
The two serve distinct purposes: expert explanation focus on model debugging, validation, refinement, and scientific understanding and \emph{popularized} explanations aim to justify decisions or build trust \cite{bhatt2020explainable, ribeiro2016should}.

\emph{Popularized} explanations are tailored for non-technical stakeholders, using simplified concepts and accessible terminology. These explanations prioritize understandability and actionability (information that enables someone to take specific, concrete actions based on that information) over technical precision (see for instance the work of Lerouge et al.~\cite{lerouge2026modeling}) As Wick et al. argue \cite{wick1992reconstructive}, these explanations must be reconstructed from technical knowledge into forms that align with recipients' mental models and prior knowledge.

By contrast, \emph{expert} explanations contain technical details and formal representations designed for developers or domain specialists who possess the necessary background knowledge to make sense of them. These explanations may include mathematical formulations, feature importance distributions, model architecture specifications, and performance metrics that would be inaccessible to lay audiences. 

The distinction between audience types influences not only the content and form of explanations but also their evaluation criteria. While \emph{popularized} explanations might be assessed on comprehensibility, persuasiveness, and usefulness for decision-making \cite{kulesza2013too}, expert explanations are typically evaluated on technical accuracy, completeness, and diagnostic utility \cite{hohman2019gamut}. This audience-centric perspective acknowledges that explanation effectiveness is inherently contextual and dependent on the recipient's goals, capabilities, and prior knowledge.

\begin{tcolorbox}[colback=green!5!white, colframe=green!5!white, breakable]
\textbf{\emph{Popularized} explanation:}  
It should avoid technical terms or parameter values (such as weights in a linear model), even when the underlying model is relatively simple. Presenting raw numerical values can confuse rather than clarify. Instead, explanations should be framed in natural, accessible language that focuses on the user’s situation.
For example: 

\vspace{5mm}

\(e =\) \emph{“Your loan application was denied mainly because your recent payment history includes a missed credit card payment and your current income is below the required threshold.”} 

\vspace{5mm}
 
However, such an explanation may be insufficiently persuasive, as users may not understand who defined the threshold, or why that particular value was chosen. A more effective and actionable explanation could be: 

\vspace{5mm}

\(e' =\) \emph{“For the actual amount of credit you are asking for, you should increase your monthly income by 100 euros to be granted the loan".} 

\vspace{5mm}
 
This second explanation is more helpful from the user’s perspective, as it provides a clear and achievable recommendation to potentially change the outcome.

\medskip
\textbf{Expert explanation:}  
In this context, more technical detail is appropriate and often necessary. For instance an explanation for auditors:

\vspace{5mm}
 
\(e =\) \emph{“The application was denied because the applicant’s risk score fell below the minimum threshold defined by regulatory guidelines. In particular, the requested credit amount exceeded 33\% of the applicant’s total indebtedness.”} 

\vspace{5mm}
 
Another explanation for model developers to understand the decision can run as follows. Suppose the ADS is a logistic regression trained on standardized features. The current decision corresponds to an instance \(x\) with the following attributes:

\[
x = \{\text{income} = 950,\ \text{age} = 28,\ \text{credit\_history} = \text{poor},\ \text{recent\_defaults} = 2\}
\]

The model outputs:

\[
f(x) = \text{deny} \quad \text{with probability} \ 0.84
\]

Feature contributions to the log-odds score are: income: \(-0.45\), age: \(-0.10\), credit\_history: \(+0.60\), and recent\_defaults: \(+0.80\).

\vspace{5mm}
 
\(e =\) ``The most influential positive factor toward denial is \texttt{recent\_defaults}, followed by \texttt{credit\_history}. Model calibration and AUC = 0.91 indicate good performance, but false negative rate remains high for applicants under 30 with low but stable income."
\end{tcolorbox}

\subsubsection{By format}

Explanations can be categorized by their representational format: \emph{visual} or \emph{textual}. Visual explanations include graphs, heatmaps, attention maps, decision trees, feature importance plots, or other graphical representations designed to convey information intuitively. For instance, saliency maps highlighting influential image regions \cite{selvaraju2017grad} and feature attribution visualizations \cite{lundberg2017unified} allow users to literally ``see'' what the model is focusing on.
 
Textual explanations, by contrast, rely on natural language descriptions, logical rules, or formal specifications to communicate the reasoning behind model decisions. These explanations range from simple feature-value statements (e.g., ``Loan denied due to insufficient income'') to elaborate narratives describing causal chains and reasoning steps. Textual explanations benefit from the precision and expressiveness of language, allowing for nuanced descriptions of model behavior and contextual factors. Rule-based textual explanations offer the additional advantage of being machine-actionable, enabling automated verification and reasoning \cite{lakkaraju2016interpretable}.

\begin{tcolorbox}[colback=green!5!white, colframe=green!5!white, breakable]
\textbf{Visual explanation:} The explanation below is a waterfall of SHAP \cite{lundberg2017unified} that explains one single prediction.
It decomposes the final prediction into the cumulative contribution of each feature, starting from a baseline value (e.g., the average model output) and showing how each feature pushes the score higher or lower. Positive contributions (in red) increase the predicted risk, while negative contributions (in blue) reduce it. 
\begin{center}
    
\begin{tikzpicture}

    \draw[very thick] (0,0) rectangle (8.3,3.2);
    
    \draw[-{Latex}, thick] (3,-0.5) -- (3,0) node[midway,right]{Base rate = 0.2};

    \fill[red!70] (3,0.5) --                     
    (5.1,0.5) --                   
    (5.4,0.9) --                   
    (5.1,1.3) --                   
    (3,1.3) --                     
    cycle;
    \node[right] at (5.4,0.9) {\footnotesize +0.4};
    \node[left] at (0,0.9) {Defaults = 2};
    
    \fill[red!70](5.4,1.3) -- (6.3,1.3) -- (6.6,1.7) -- (6.3,2.1) --    (5.4,2.1) -- cycle;    
    \node[right] at (6.6,1.7) {\footnotesize +0.25};
    \node[left] at (0,1.7) {Income = 900};
    
    \fill[blue!60](6.6,2.2) -- (6,2.2) -- (5.7,2.6) -- (6,3) -- (6.6,3) -- cycle;
    \node[left] at (5.4,2.5) {\footnotesize -0.1};
    \node[left] at (0,2.5) {Job = stable};
    
    \draw[-{Latex}, thick] (5.7,2.9) -- (5.7,3.8) node[right]{Output = 0.75};

\end{tikzpicture}
\end{center}

\textbf{Rule-based textual explanation:}  \\
\(e =\) \emph{“income $< 1000 \wedge \text{ recent\_defaults }> 1 \implies \text{score} > 0.6 \implies \text{deny}$ ”}  

This form uses a logical rule either derived from the model’s structure, or approximated from its input-output behavior using techniques such as rule induction.
\vspace{1em}

\textbf{Natural language narrative:}  
\(e =\) \emph{“Your loan application was denied because your reported monthly income of 900 euros is below the minimum required threshold of 1200 euros, and your credit history indicates two recent defaults. These factors suggest a high risk of non-repayment, which exceeds our institution’s acceptable risk level.”}  

\end{tcolorbox}

\subsubsection{By formal reasoning}

When generating an explanation, we are implicitly answering a specific type of question such as ``Why did this happen?'', ``Why outcome P instead of Q?'', or ``What would happen if...?''. Each of these questions reflects a distinct form of underlying reasoning, which guides how the explanation is constructed and interpreted. Thus, one meaningful way to categorize explanations is by the type of formal reasoning they embody.

\emph{Contrastive} explanations address why-not questions of the form ``why P rather than Q,'' where P is the actual event that occurred and Q is a \emph{counterfactual} contrast case that did not occur (a hypothetical alternative outcome). As Miller \cite{miller2019explanation} demonstrates, this form of reasoning is particularly powerful because humans naturally think in terms of contrasts rather than absolute explanations. These explanations focus on the differences between the actual outcome and the expected or desired alternative, highlighting causal factors that differentiate P from Q.

\emph{Counterfactual} explanations respond to how-to questions and present hypothetical \textit{alternative inputs} that would lead to different outcomes \cite{korikov2021counterfactual, wachter2017counterfactual}. Unlike contrastive explanations, which focus on the outcome, counterfactuals emphasize manipulable features or conditions that would alter the result. These explanations are particularly useful in actionable scenarios where stakeholders need guidance on how to achieve different results. For example, in a loan application scenario, a counterfactual might specify the minimum income needed for approval, whereas a contrastive explanation would explain why the application was rejected when compared to successful applications.

\emph{Abductive} reasoning forms another foundation for explanations, representing inference to the best explanation \cite{harman1965inference, leake1995abduction}. This approach begins with observations and works backward to determine the most likely cause or explanation that accounts for those observations. Abductive reasoning is particularly valuable in complex systems where multiple factors may contribute to an outcome, and the most plausible explanation must be identified from competing possibilities. This form of reasoning often involves evaluating multiple hypotheses against criteria like simplicity, coherence with background knowledge, and explanatory power.

\emph{Causal} reasoning underlies explanations that identify genuine cause-effect relationships rather than mere correlations. These explanations answer what-if questions by mapping the causal structure that connects inputs to outcomes. Drawing from Pearl's causal calculus \cite{pearl2009causality}, causal explanations distinguish between direct effects, indirect effects, and spurious associations. This type of reasoning allows for more robust explanations that can anticipate system behavior under novel conditions or interventions, making it particularly valuable for critical applications where understanding the underlying mechanisms is essential.

\begin{tcolorbox}[colback=green!5!white, colframe=green!5!white, breakable]

\textbf{Contrastive explanation:}  
\emph{“Your loan application was denied (P) rather than approved (Q) because your reported income was 1000, whereas applicants who were approved had incomes above 1500, and you have two recent defaults while successful applicants had none.”}  

\vspace{1em}

\textbf{Counterfactual explanation:}  
\emph{“If your monthly income had been at least 1200, and you had no defaults in the past year, your loan would have been approved.”}  

\vspace{1em}

\textbf{Abductive explanation:}  
\emph{“The loan was denied likely because of a combination of low income, multiple recent defaults, and poor credit history. Among these, recent defaults appear to be the strongest contributing factor.”} 

\vspace{1em}

\textbf{Causal explanation:}  
\emph{“Your loan was denied because your income directly influences the risk score, and income below 1200 causes the score to fall below the approval threshold. Other factors like age and employment have indirect or negligible effects in this case.”}  

\end{tcolorbox}

\subsubsection{By discursive interaction}
Another body of work move away from viewing explanations as static elements and instead present them as \emph{processes} that unfold through interaction. 
As Miller \cite{miller2019explanation} argues, explanations should be understood as forms of social interaction rather than as static pieces of information. In this perspective, an explanation is not merely a statement of causes or reasons but a \emph{conversation} between an explainer and an explainee. Drawing on Hilton’s \cite{hilton1990conversational} conversational model of explanation, Miller highlights that explanations, like conversations, are governed by cooperative principles. They must address the explainee’s question, remain relevant to the context, and adhere to Grice’s conversational maxims of quality, quantity, relation, and manner \cite{grice1975logic}. These maxims imply that a good explanation should be truthful and supported by evidence (quality), sufficiently informative without unnecessary detail (quantity), pertinent to the recipient’s concern (relation), and expressed clearly and coherently (manner).

Closely related to the dialogue perspective, the dialectical approach adds a layer of \emph{argumentation}. While dialogue aims at mutual understanding through cooperative exchange, dialectic seeks to challenge and justify positions through reasoned debate. It thus goes beyond a simple question–answer model, aiming not merely to report causes but to support claims with arguments \cite{walton2006examination, walton2011dialogue}.
Walton \cite{walton2011dialogue} propose dialogical models to implement Hilton's conversational model, composed of three stages: an opening stage (where the need for an explanation is established), an exploration stage (where reasons are exchanged), and a closing stage (where understanding is assessed). This model addresses a central issue: how to determine when an explanation is complete. Walton \cite{walton2006examination} argues that understanding is not achieved merely when information is transmitted, but when the explainee can demonstrate comprehension by answering new, related questions. Moreover, selecting the \emph{best} explanation does not mean the best one possible (as in abductive reasoning), but rather identifying the explanation that is arguably stronger than its alternatives. Because explanations rely on \emph{defeasible} reasoning ---that is, reasoning open to revision or challenge --- they are naturally framed within argumentation schemes such as arguments from analogy, from expert opinion and so on \cite{walton2007dialogical}.

In a socioeconomic perspective, scholars from the ``theory of conventions" put the emphasis on the fact that in the economic and social world, there is a variety of frameworks of interactions, named by Boltanski and Thévenot \cite{boltanski2006justification} the ``cités''. For instance, in the ``market cité'', the motive for action is the profit. In the ``industrial cité'', the cooperation between workers, bosses and machines is guided by the search for productive efficiency. In the ``civic cité'', what is valued are the democratic principles or the social commitment of citizens. To sum-up what appears valuable in this theory, in each cité a supreme principle provides a shared communicative pattern based both on reciprocal expectations. The common supreme principle is not intangible: it is continuously subject to criticism. Boltanski and Thévenot call this process ``justification''. The principles of profit in the market, of technical efficiency in the industry, of social commitment in the community can be criticized and must prove their legitimacy. From this socio-economic theory, we must retain the ideas that the justification operation presupposes a common framework, and that it must be appropriate to the context in which it is deployed.

Hénin and Le Métayer \cite{henin2022beyond} also emphasise the importance of justification and contestability. They provide clear definitions of each concept and explicitly distinguish between explanation and justification, as we also aim to do in this paper (in subsection~\ref{expvsjust}, we further clarify how our approach differs from theirs). Their work is also situated within this typology of explanatory models, since contestability introduces a dialectical dimension and they do it with reference to norms. In their framework, called \emph{Algocate} \cite{henin2021framework}, users can interact directly with the decision system by asking questions and receiving tailored explanations, thereby transforming explanation into a cooperative conversational process. The third example below is inspired by their framework.

\begin{tcolorbox}[colback=green!5!white, colframe=green!5!white, breakable]
\textbf{Dialogical explanation}  
Suppose an applicant’s loan request is denied and the decision-maker provides an interactive interface that allows the applicant to ask questions and receive contextually adapted explanations. This interactive approach operationalizes the conversational nature of explanation proposed by Hilton \cite{hilton1990conversational}. 
\begin{itemize}
    \item[\emph{User}] ``Why was my loan denied, even though my revenue is greater than $5000$ and my credit amount is lower than $2000$” 
    \item[\emph{System}] ``Your recent credit history shows two missed payments, which increased your estimated risk of default.” 
    \item[\emph{User}] ``What can I change to be accepted?”  
    \item[\emph{System}] ``If your average income over the last six months were $200$ higher, your risk score would fall below the approval threshold.”

\end{itemize}

\textbf{Dialectical explanation}  
Here, the user can challenge the reasons offered, question the decision’s legitimacy and the model’s assumptions. Explanation thus evolves into justification through argumentation, reflecting a shift from understanding to critical deliberation. 
\begin{itemize}
    \item[\emph{User}] ``Why was my loan denied even though my income is stable?”

    \item[\emph{System}] ``Your loan application was denied because, according to financial experts, applicants with more than two recent defaults are highly likely to default again. Our system follows this principle to minimize risk.” (\emph{argument by expert opinion})
    \item[\emph{User}]  ``I understand, but my defaults were due to a short-term health issue. Since then, I have repaid all my debts on time. Shouldn’t my recent repayment history count more than older defaults?"
    \item[\emph{System}] ``The rule we apply generalizes from expert knowledge, but your recent financial stability may indicate that this generalization does not hold in your case. We can review your situation in light of this new evidence."
\end{itemize}

\textbf{Norm-Based justification in the sense of Hénin and Le Métayer: }suppose  an applicant contests the decision so the decision-maker has to justify it.
\begin{itemize}
    \item[\emph{DM}] Your loan application was denied because your debt-to-income ratio exceeds 45\%. According to our institution’s credit policy, based on the national financial regulation ``Code du crédit à la consommation'', loans cannot be granted above this threshold.
    \item[\emph{User}] But I’ve recently obtained a permanent job that increases my income. Shouldn’t that be taken into account before applying the rule?
    \item[\emph{DM}] The regulation aims to prevent over-indebtedness and ensure borrowers’ financial stability. Even though your situation may improve, we must apply the same standard to all applicants to remain compliant with this fairness principle.
\end{itemize}

This dialogue illustrates a \emph{norm-based justification}, in which the decision-maker refers explicitly to a regulatory norm to legitimize the outcome. In contrast to mere \emph{explanation}, which describes causal mechanisms, \emph{justification} appeals to shared principles (legality or procedural consistency). The exchange also reflects Hénin and Le Métayer’s idea that justification connects the decision to the normative framework governing the system.
\end{tcolorbox}

\begin{figure}
    \centering
    \includegraphics[width=\textwidth]{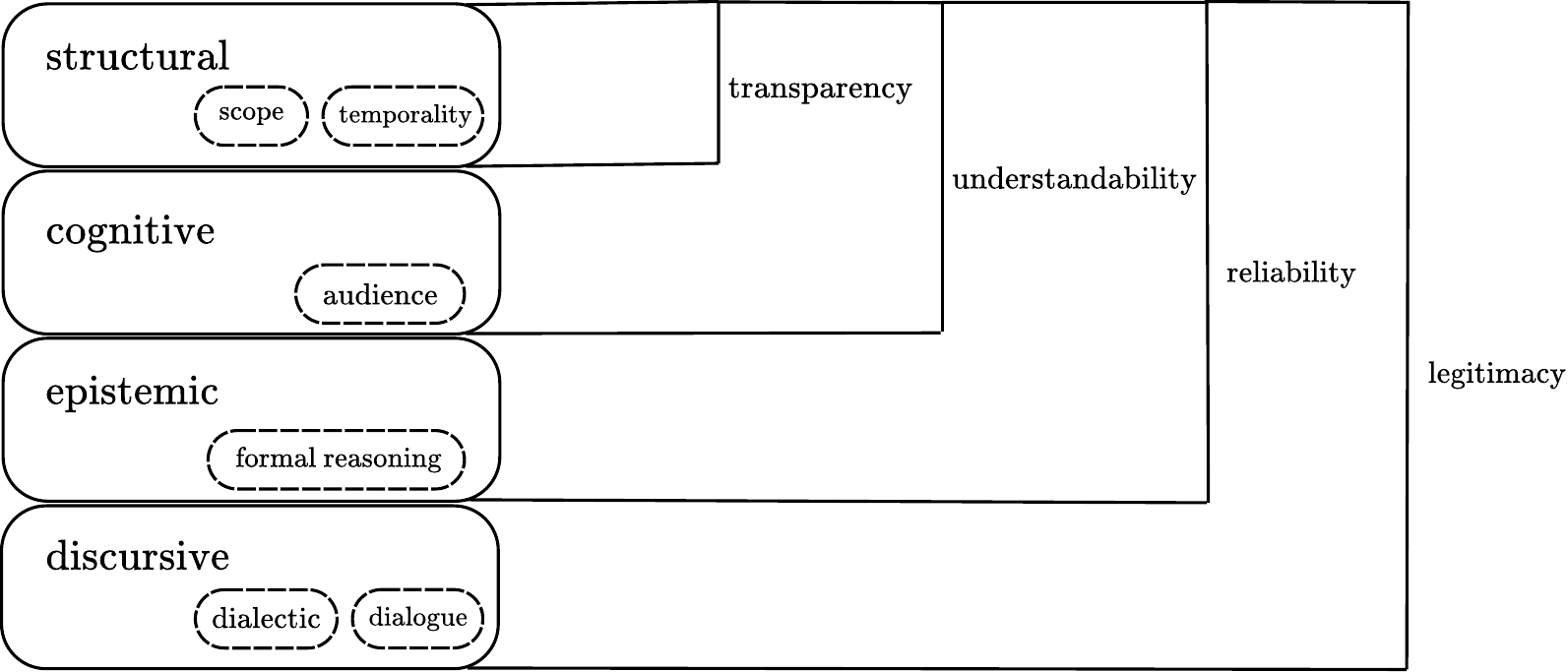}
    \caption{Conceptual bridge between explanation typologies}
    \label{fig:typologies}
\end{figure}
To conclude, figure \ref{fig:typologies} summarizes the conceptual evolution of explanatory approaches in the literature. Early work focused primarily on the \emph{structural} dimension of explanation, emphasizing model interpretability through mechanisms such as feature importance or visualization. This was followed by \emph{cognitive} approaches, which aim to align explanations with human reasoning processes and cognitive biases, as discussed by Miller \cite{miller2019explanation}. A further shift introduced \emph{epistemic} perspectives, emphasizing the explanatory value of causal knowledge and truthfulness in model reasoning. Finally, other work extends toward \emph{normative} approaches, which view explanation as a discursive process grounded in social interaction and argumentation. These notions can be understood as forming a hierarchy of inclusion. Transparency represents the most basic level, alone it is insufficient if the information provided cannot be meaningfully interpreted, hence the need for comprehension, which presupposes transparency but adds information that aligns with the expectations and mental models of the recipients. Reliability builds upon comprehension, since users can only develop trust in a system they understand and perceive as consistent in its behavior. Finally, legitimacy encompasses all previous levels: a system can be transparent, understandable, and reliable, yet still fail to be legitimate if its outcomes are not perceived as such or aligned with shared norms.

Our work is situated within the discursive category, and builds upon it.
While prior argumentative approaches succeed in capturing the interactional dimension of explanation, we argue that they do not ensure \emph{acceptability}. Achieving \emph{acceptability} requires more than legitimate process: it requires that the decisions be justified in the eyes of their recipients. To articulate this distinction precisely, we draw on Habermas’s theory of communicative action, which provides a framework for evaluating the rationality of communicative exchanges, and on Perelman’s theory of argumentation, which grounds the success of justification in the recipient's reasoned adhesion. On this basis, we propose a clear conceptual distinction between \emph{explanation} and \emph{justification}. This distinction is not merely terminological: it identifies a gap in existing approaches and motivates a new conceptual framework, developed in Section \ref{section3}, designed to bring algorithmic decision-making systems closer to collect acceptability.

\section{A conceptual framework}
\label{section3}

The former section explored the vast, multidisciplinary literature on explanation, explainability and justification. This review highlighted the fact that this literature presents different typologies of concepts that are all to some extent relevant, but stem from different logics and hence explore different aspects of the diversity of concepts, issues and ideas surrounding explanation, explainability and justification. Taken one by one, the different typologies appear focus on specific aspects of their subject matter, but an overarching conceptual framework encompassing the specific strengths of these various approaches and building upon their complementarities is found lacking in the literature. Our point in this section is to propose such a conceptual framework. 

The crux of our proposal is to conceptualize explanations and justifications as unified by their shared anchorage in argumentation and social exchanges mediated by language, but distinguished by the specific roles they play in different kinds of discussions with different types of recipients.

The shared anchorage we propose refers to Meinard and Tsoukiàs \cite{meinard2019rationality}, who argue that the philosophy of Jürgen Habermas can usefully shed light on the analysis of forms of rationality in decision support theory and practice. They introduce an analytical framework leading to a Habermas-inspired typology of decision aiding approaches. Their typology, which distinguishes between ``objectivist, conformist, adjustive and reflexive approaches'', is intended to be useful to practitioners, who should be able to identify which approach they should use in a given situation. Below, we show how this framework can be used as a common frame for both explanation and justification, and how it manages to encompass the strengths of the various typologies explored in the former section.

Within this common frame, we propose the following clarification of the distinctive roles of explanation and justification. Whereas explanations serve to clarify the decision-making process, justifications are designed to ensure that the decision will be perceived as convincing and legitimate. This aligns with Perelman's concept of argumentation, where the success of the exchange is determined by its capacity to gain the recipient's acceptance, which is achieved through a combination of clear explanations and persuasive justifications \cite{perelman1958traite}. 

For the purpose of developing and presenting this conceptual proposal, we make some working simplifications. Although automated decision-making systems usually involve multiple stakeholders, with multifarious roles, rights, duties and stakes, such as the decision-maker, the system-designer, the lay recipient of the decision, the explanation and/or justification issuer, the recipient of such explanation and/or justification...

In what follows, we essentially focus on who issue the explanation and/or the justification (denoted IS) and who receive them (denoted RCP) since our framework is built upon their interaction. 
As a further simplification, regarding the algorithm, we do not differentiate between white-box and black-box models, as our focus is on the presence of an algorithm ---regardless of its nature and internal complexity---in the decision-making process. In both cases, the need for explainability remains the same. However, the type of explanation required differs. White-box models may offer inherently interpretable explanations due to their simple structure, whereas black-box models often require post-hoc methods to provide meaningful insights into their decision-making process.

\subsection{Explanation vs. Justification}
\label{expvsjust}

Let us define an \textbf{explanation} as the operation that consists in giving the characteristics of the process that led to the decision. In other words, an explanation consists in informing the RCP (whether it is the decision-maker or the recipient of the decision) about what actually happened in the decision-making process, factually and historically. The main goal of an explanation is that its recipient should \emph{understand}.

As opposed to explanations, a \textbf{justification} goes beyond factual recounting (the hallmark of explanations) to assess whether the decision was normatively acceptable. Justifications offer reasons aimed at establishing the legitimacy of a decision, meaning reasons that the affected individual can find reasonable. 

Both explanations and justifications are shaped by practice, as they primarily concern an action or a disposition to act \cite{perelman1961legal}, and they are both typically context-dependent. Because the context is not always immutable, various elements of both explanations and justifications can be subject to reconsideration if compelling reasons are presented to challenge them \cite{perelman1961legal}. In Habermasian terms, both explanations and justifications are situated in the ``life-world” \cite{habermas1987com}, which implies that they must resonate with people's everyday understanding and shared social norms. In justifications, this reliance on everyday norms tends to overweigh the reliance on technical details, while the later can play a more important role in explanations. However, both elements can have a role to play in both explanations and justifications.



Other scholars have already given significant weight to the difference between explanation and justification. For example, for Hénin and Le Métayer \cite{henin2022beyond}, the distinction between these notions is relevant because they have specific properties and goals. They consider ``Explanations (as) transfers of knowledge (from the ADS to the explainee)'' that are descriptive and intrinsic; in contrast, justifications are normative and extrinsic in the sense that the judgment about the goodness of an outcome must be based on an external reference such as a legal or ethical norm. Although we concur with Hénin and Le Métayer that explanations and justifications should be distinguished, we depart from them on two important ideas. First, the reference to an external norm is not a necessary condition for a justification; second, a justification is connected to a search of adhesion, in an argumentative process. These ideas will be further elaborated in the next sub-section in which the explanation/justification issue will be framed with reference to a typology of explanation models.

\begin{tcolorbox}[colback=green!5!white, colframe=green!5!white, breakable] Same context as \ref{extyp}. Suppose we have $f(x) = s \geq t$ for individual $x$, the DM provide an explanation to the applicant : 

$e_{DM} = $ \emph{``the loan was denied because the current income is below the threshold required to safely cover monthly loan installments''.}

In this case, if the explanation process ends here, the applicant may question the legitimacy of the threshold and find the explanation unacceptable.

A justification could be:

$j_{DM }= $\emph{``The decision was made in accordance with the institution’s credit policy, which is designed to ensure applicants are not placed at financial risk they cannot bear. The income threshold is set based on statistical models and regulatory guidelines that balance financial inclusion with responsible lending. In cases where applicants are near the threshold, we offer alternative financial products or re-evaluation after updated income data.''}

This constitutes a justification because it provides normative reasoning and offer reasons that a reasonable person could accept, even if they disagree with the outcome.
\end{tcolorbox}

\subsection{A typology of explanation and justification models}

We propose a typology which distinguishes four explanatory models. It is build mainly upon Habermas' models of action and, in a specific model, upon  Perelman's theory of argumentation. In what follows, the models we propose are \emph{ideal-typus} of explanation and justification of decisions based on an ADM. We will argue that the four models that will be proposed do not perform equally in terms of justification, adherence and finally legitimacy. 

\subsubsection{Technical model}
\label{tech_model}

This first type is based on Habermas' strategic model of action, which accounts for an agent's action in terms of objective facts \cite{meinard2019rationality}. Here, the algorithmic model is considered rational owing to its solid scientific and technical grounds. A technical explanation is hence exclusively based on scientific and technical considerations. A technical explanation can be used to explain the functioning of a given algorithm to the decision issuer. If used to convince a lay recipient that this algorithm produced legitimate outcomes, or functions in a legitimate way, then in this usage it plays the role of a justification. Using technical explanations in this way can foreseeably have inconvenient implications:

\begin{itemize}
    \item It is likely that there will be a gap in understanding between IS and RCP.
    \item For the RCP, the justification provided can fail to be convincing, and the legitimacy can hence be contested.
\end{itemize}

\subsubsection{Norm-oriented model} 
\label{norm_model}

In this model, both explanations and justifications are based on compliance with legal or moral standards (such as non-discrimination, equal treatment, mandatory human intervention to issue the final decision...). An explanation is norm-oriented when it accounts for the objective, historical fact that this or that choice was made, when elaborating the algorithm, \emph{in order to} comply of the norm at issue. A norm oriented justification argues that the functioning or the outcome of the algorithmic system is legitimate \emph{because} it complies with this norm, whether this compliance was intended beforehand or not.

The following corollaries are worth mentioning:
\begin{itemize}
    \item A norm-oriented explanation formalizes and explains the specific norms that were used when elaborating the algorithmic model, their interpretation, and the way they are implemented in the model.
    \item Legal norms and subjective fairness do not necessarily overlap. Legality or alignment with social norms certainly are sources of legitimacy, but if a given decision is substantially disadvantageous for a given RCP, the general justifiability bestowed by legality or social norms can be outweighed by personal disadvantage in the RCP's overall assessment. In such cases, norm-oriented justifications can require, in order to be successful, detailed and convincing justifications of why a given (generally acceptable) norm has had this or that particular implication for the RCP.
    \item As opposed to norm-oriented explanations, a norm-oriented justification may appeal to a norm that did not play any role in the design and motivation of the algorithm, if the functioning and/or outcome appear to abide by this norm. This approach can feed more powerful justifications, if the norm at issue is more widely shared and hence proves more convincing for the RCP.
\end{itemize}

\subsubsection{Expressive model} 
\label{expressive_model}
The hallmark of the expressive model is the idea that rationality stems from the sincere expression of motives, preferences and values. An expressive explanation will hence faithfully track and report the motives, preferences and values that the designer of the algorithmic model has (possible unconsciously) expressed through the choices s/he has made when elaborating the algorithmic model. An expressive justification will herald motives, preferences and values that its recipient can recognize as his/her own.

\subsubsection{Communicative model} 
\label{com_model}

The fourth model echoes Habermas's ``communicative rationality'', which involves impartial discussions and the search for the best argument. Its implementation implies a communicative activity. The model is also consistent with Perelman's theory of argumentation, which stresses the following ideas: argumentation essentially is a search for convincing argument; it unfolds as a progressive process; it must be coherent with the recipient's needs and expectations; the acceptability of an argument is context-dependent; there is hence no such thing as an argument that would be totally binding in itself.

Explanation and justification, in the communicative model, hence involve a communicative activity that consists in producing and exchanging arguments. When this communicative activity serves the purpose of fine-tuning the account given of how the algorithmic system functions, in order to foster the recipient's understanding, then this communicative activity yields a communicative explanation. If it rather aims at convincing that decisions or recommandations are legitimate or well-founded, then it constitutes a communicative justification.

This model is largely, though not entirely, consistent with Miller's framework \cite{miller2019explanation}, which is itself inspired by Hilton's conversational model \cite{hilton1990explanation} and Walton's dialogical theory of explanation \cite{Walton2011explanation}. Our model shares with these alternative frameworks an emphasis on conversation and dialogue between explainer and recipient.

\subsubsection{Summary}
Table \ref{tab-conceptual} synthesises the conceptual framework developed above. A key observation is that the same model serves different purposes depending on whether it is used as an explanation or a justification, and that these purposes do not always align.

A \emph{technical} explanation is a correct recount of how the decision was made; a \emph{technical} justification shows how the decision complies with standards of scientific accuracy. A \emph{norm-oriented} explanation accounts for the norms that actually motivated the design of the system; a \emph{norm-oriented} justification reflects more the issuer's policy choices than the norms that were operationalised in practice.
An \emph{expressive} explanation faithfully tracks the motives and values expressed through the designer's choices; an \emph{expressive} justification appeals to motives and values the recipient can recognise as their own. Whether these two sets of values coincide is an empirical question, and when they do not, expressive justification will fail.

A \emph{communicative} explanation deploys argumentation to foster the recipient's understanding of how the system functions; a \emph{communicative} justification deploys the same argumentative resources to establish that the decision is legitimate and well-founded. The unifying activity — the exchange and critical assessment of arguments — is the same in both cases. What differs is the goal.
Explanation succeeds when the recipient finds the account acceptable, that is, intelligible, relevant, and consistent with their expectations and mental models. Justification succeeds when the recipient finds the decision legitimate, that is, when the reasons offered can be endorsed through reasoned deliberation, even in the face of a personally unfavourable outcome.

\begin{table}[h]
\centering
\resizebox{\textwidth}{!}{%
\begin{tabular}{|l|c|c|}
\hline
\textbf{Model} & \textbf{Explanation} & \textbf{Justification} \\ \hline

Technical 
& \begin{tabular}[c]{@{}c@{}}Scientific accuracy\\ Correctness\end{tabular} 
& Compliance with technical standards \\ \hline

Norm-oriented 
& Compliance with existing norms 
& Policy / rationality oriented \\ \hline

Expressive 
& Faithfulness 
& Meeting the recipient value system \\ \hline

\multirow{2}{*}{Communicative} 
& \multicolumn{2}{c|}{Argumentation} \\ \cline{2-3}

& Acceptability 
& Legitimacy \\ \hline

\end{tabular}%
}
\caption{Conceptual framework of explanations and justifications}
\label{tab-conceptual}
\end{table}

\subsection{Implications of the overall structure of the conceptual framework}

By definition, one cannot expect a justification to be successful if its issuer does not share some common ground with its recipient. Sharing the same model appears to be a minimal requirement in this regard. The choice of a model that the recipient will endorse is hence crucial for any issuer of a justification. When making this choice, the issuer should be aware that the communicative model (model \ref{com_model}) has distinctive strengths when the interests of IS and RCP diverge. 


Notice that the justification given by the IS may be strategic: it may indicate reasons deemed acceptable to the RCP, instead of the true reasons. For instance, consider a company that uses an automated system to filter job applicants based on their residential address, systematically rejecting candidates who live outside a specific, affluent neighborhood. On the surface, the IS might offer the following justification: \emph{``We prioritize hiring people who live nearby in order to support the local economy and strengthen community ties.''}
While this justification may appear socially responsible, it can mask underlying discriminatory motives, such as using geographic location as a proxy to exclude candidates from less wealthy or more ethnically diverse areas. In this case, the justification functions more as a socially acceptable narrative than as a sincere account of the decision's normative legitimacy.


\section{Case study: admission in French universities}
\label{section4}

In this last section, we develop the application of our conceptual framework in a case study: the process used in France for university admission at two levels: undergraduate (first-year) studies and Master’s programs. We analyze this empirical case through the lens of the above conceptual framework and use the later to assess the current system in an original perspective. Based on these findings, we draw lessons that might prove useful for both developers and decision-makers in their efforts to improve the way rebuttals are justified to students.

\subsection{Properties and functioning of the system}

Access to universities in France has undergone radical change since the late 2000s. Traditionally, the right to higher education has been a prominent principle of the French system. Unlike so-called ``grandes écoles'', where admission is decided by highly selective entrance examinations, admission to university has traditionally been open to all. However, this principle of a ``right to higher education" was challenged in recent years by the use of automated algorithms for selecting students applying to different university disciplinary programs. After an initial experiment in the ``Ile-de-France'' Region in the 1990s, the State introduced a national application in 2009: APB (``Admission Post-Bac''), which was replaced by a new algorithm in 2018, Parcoursup, which is currently still in use.

Under the APB system, high school students in their final year had to prioritize wishes (with a maximum of 24 wishes). Assignments were made using local algorithms determined by the universities. APB was plagued by numerous malfunctions in high-volume disciplines such as medicine, law, psychology and sport, in which lotteries had often to be used to decide between numerous candidates of equal merit. This practice lacked a legal basis, as stressed by the highest French administrative court - the ``Conseil d'Etat''  -  in a decision dated December 22, 2017. The Ministry of National Education has subsequently replaced APB with Parcoursup, which has been legally in force since March 8, 2018 (``loi relative à l'orientation et à la réussite des étudiants'', known as ``the ORE law''). Parcoursup is based on the ``stable matching problem'', solved using the Gale-Shapley algorithm \cite{Gale1962algorithm}. In this process, students in their final year of high school submit 10 non-hierarchical wishes for each academic program, which can be broken down into 10 sub-wishes for each university. 

The procedure comprises 3 phases, with a very tight timetable: the platform opens in November, at which point students fill in a “dialogue form” for the Class Council's opinion; from January to March, applicants formulate their wishes; then, from May to July, the universities' responses are communicated, and students make their final choice. Parcoursup concerns access to undergraduate studies. Access to post-graduate courses is orchestrated by a dedicated platform: Monmaster. In principle, universities must provide as much information as possible about their selection and admission criteria, as well as their intake capacity for 1st year students in the various courses, specifying the number of places reserved for students who benefit from social scholarships. If rejected or placed on a waiting list, candidates have two options. The first option is a hierarchical appeal. Such appeals are handled by the ``Commission d'Accès à l'Enseignement Supérieur'' (CAES), under the responsibility of the ``Recteur'' (the regional head of the education administration). There are around 23,000 appeals per year (an apparently stable figure: 23000 in 2018, 25000 in 2019, 23400 in 2023). The alternative option is to submit a discretionary appeal. Such appeals are submitted either to heads of courses (usually by e-mail), and then consist in a request to be admitted to a given first year program or to a given Master's course, or to President of Universities (by registered letter), in which case they inquire about the reasons motivating rebuttals. There are between 500 and 1,000 appeals of that sort per year in a specific department" which ``can lead universities to process appeals automatically, generating standardized responses which are calibrated in advance by the legal and academic departments” (Allouch and Espagnolo-Abadie,p. 21 \cite{allouche2024contest}).

Although Parcoursup is a national platform, it is implemented at a local scale by universities, which are autonomous since a reform initiated in 2007. This leads to a wide variety of academic assessment criteria, depending on the university, or even on the department within a given university, which is in charge of applying these criteria. Moreover, the context matters: not all universities operate under the same logic when it comes to student selection. Some institutions have a long-standing elitist culture, and for them, the introduction of student selection via Parcoursup is not a break with the past - quite the contrary. Others are still attached to the right of all ``baccalauréat'' holders to access higher education; they have reluctantly switched to a selective system and are critical of the selection process, which they claim disadvantage underprivileged social classes. A third group of universities combines low selectivity for first-year admissions with high selectivity at the Master’s level. In this context, appeals against first-year non-admissions mostly arise in the second category of institutions, while appeals concerning Master’s-level rejections are primarily seen in the third.

The variety of academic assessment criteria mentioned earlier is perceived as a source of opacity by applicants to the Parcoursup algorithm, as underlined by a 2020 parliamentary report \cite{Juanico-Sarles2020report}. This perceived opacity is reinforced by the lack of information on the selection methods used in practice: some universities select ``by hand'', while others use local algorithms. 

According to a 2023 IFOP (a pollster) survey, 83 percent of respondents find the Parcoursup experience stressful. A 2019 survey by the ``Observatoire de la Vie Etudiante'' showed that 27 percent of students find the Parcoursup platform unfair \cite{Belghith2019orientation}. Indeed, appeals against non-admission decisions are based on a feeling of injustice, particularly when students have a very good academic record and adhere (along with their families) to the meritocratic system, and all the more so when the parents have higher education qualifications, belong to the upper classes and are attentive to their child's studies.

\subsection{Assessment: does the system manage to successfully justify rebuttals?}

The problem of justification of rebuttals is a sensitive one. Candidates whose demands have been rejected are very often (if not always) given standardized explanations, sometimes automatically generated by local algorithms. When program managers receive 500 to 1,000 requests in a short period, all asking for the reasons behind their rejection, they lack both time and human resources to respond adequately. Moreover, professors are generally not accustomed to this kind of exercise. There clearly is a discrepancy between the expectations of rebutted candidates (and their families) -- particularly when applicants were very good high school students, invested in their education.W hen confronted with a non-admission decision in their most preferred academic programs, they feel disoriented or destabilized by both the negative decision itself and the nature of explanations eventually provided. University professors are aware of this tension but often feel have no viable alternative \cite{allouche2024contest}.

The uncomfortable position of university professors has been documented by
Allouche and Espagnolo-Abadie \cite{allouche2024contest}. They quote the head of a psychology department expressing discomfort about the selection process, arguing that  relying simply on grades instead of applicants' motivation is ``more practical and more protective for teaching teams''. This professor also  admits that, given the limited number of available spots, very good candidates may end up being rebutted. These empirical elements can be integrated into the typology of explanation and justification models defined above (in section \ref{section3}).

\begin{itemize}
\item Technical Model: the IS (in this case, the Ministry of Education) has developed a Gale-Shapley algorithm that is supposed to be fairer and technically more advanced than the previous algorithm (``Admission Post-Bac'' - APB). Technical assistance is provided for users. However, this technical assistance is not what rebutted applicants expect. Indeed, the later expect assistance in terms of guidance advice, answering questions such as: is it appropriate to include this or that course or university in their wishes? The assistance device fails to provide candidates with relevant answers to such questions, and rather focuses on algorithm design. Therefore, what the assistance provides is not seen as an \emph{explanation} by users, since it does not focus on the aspects of the system's functioning that those users are interested in; \emph{a fortiori} it does not provide a convincing \emph{justification}.

\item Norm-oriented model: Parcoursup designers consider the algorithm to be fully compliant with legal norms, including the law that mandated its use (the ORE law), the GDPR and relevant European regulations that French law must comply with. In other words, the IS asserts that the system operates within the bounds of legal norms. However, rejected applicants are not primarily contesting the legality of Parcoursup. Rather, they challenge the gap between the decisions it produces and pervasive meritocratic social norms, namely the belief that a good student deserves access to the university of his or her choice to pursue the studies of their choice. Thus, compliance with legal norms from the IS's perspective may generate a disappointed meritocracy. In this case, two kinds of norms are at odds: legal and social norms. The argument of legality provides a kind of justification which confronts the wish of a meritocratic-based admission in university. Available norm-oriented justifications hence fail because the norms they refer to, which can indeed by the ones that a norm-oriented explanation should refer to, are not the norms that applicants would accept as structuring in convincing justification.

\item Expressive model: rejected applicants often criticize the dehumanization of university admissions and the impossibility of direct contact with course managers or university decision makers. Meanwhile, the IS considers the algorithm as a highly efficient matching system between applicants' preferences and academic programs capacities. The reasons given to rejected applicants are standardized and fail to meet their expectations, particularly for those who believe in selection based on merit. Allouche and Espagno-Abadie \cite{allouche2024contest} (p. 110 to 113) cite parents' criticisms of the system: some consider that everyone hides behind the Parcoursup algorithm, others that no convincing explanations are given and even that dehumanization leads to their child's value being discredited . Therefore we can conclude that the provision of explanations about the properties of the algorithm fail to satisfy the demand of justifications expressed by the candidates and their family. 

\item Communicative model: rejected applicants tend to express two types of expectations: first, to have direct human interaction with academic program coordinators to try to convince them that they deserve to be admitted, and second, the need to understand the reasons behind their rejection. Most of the time, they request both an explanation for the decision and a justification. These expectations are often unmet due to the absence of opportunities for meaningful communication or dialogue with the university. Furthermore, the actual reasons behind admission refusals are frequently concealed. For example, according to Allouche and Espagnolo-Abadie \cite{allouche2024contest}, regulations prohibit selecting Master's candidates based on grades. Yet, according to one Master’s program director in law cited by the authors \cite{allouche2024contest}, selection in practice is indeed grade-based. The Monmaster platform then records a “satisfactory level”, which is supposed to “justify” the refusal, when those admitted have received an assessment of “excellent level”. The two authors cite the case of a student applying for a Master 2 in notarial law. She had a very respectable academic record, but was rejected. She felt that her career plan, which was to become a notary, her motivations and the fact that she had completed internships in the profession, had not been taken into account. In other terms, the communicative model is able to provide a convincing justification only in the case where a genuine dialogue takes place between the decision issuer and the recipient. All in all, the system as it is currently used hences does not produced any communicative explanation or justification.

\end{itemize}

The analysis above shows that, although the system as it stands involves some attempts at explaining and justifying decisions made, by and large, these attempts fail. As highlighted in our discussion of explanations anchored in the technical model, such tentative explanations fail because they fail to grasp the expectations of their recipients. Similarly, as shown in our discussion of norm-oriented attempted justifications, available justifications misidentify the norms that recipients are liable to adhere to, and tend to mistake a norm-oriented explanation for a norm-oriented justification. A thorough application of our framework in the very process of producing explanations and justifications would help avoid such mistakes, and thereby help design more satisfactory explanations and more convincing justifications.

That being said, as highlighted in our discussion of an absence of any communicative approach in the system as it stands, because the decisions made through this algorithm are highly sensitive and because the context is very complex in most cases, it seems clear that applications of the communicational model are the most promising. The challenge, for designers and champions of this system, is hence to develop technical means to enable an argumentative dialog between issuers and recipients of explanations and justifications. 

To be sure, such perspectives are bound to be hampered by the increasing shortage of human and material resources in French universities. But that is another story.

\section{Conclusions}

Contemporary societies are unmistakably characterized by a growing pervasiveness of decision-support and recommendation systems, mostly algorithmic in their structure, spanning all the domains of our ordinary live (from everyday consumption to the organization of professional live, through leisure activities, medical care and education) as well as exceptional events such as natural disasters or financial crises.

The pervasive use and very existence of these algorithmic systems have profound, and often unintended and/or unanticipated effects on the states and dynamics of numerous aspects of our societies, including inequalities in income, life conditions and life prospects of various people and communities. Accordingly, there is a growing awareness that these systems should abide by certain norms and expectations, which have historically been applied to human agent rather than to algorithmic systems, and therefore require being adapted to these objects.

Exploring the literature addressing this pressing need, we have shown that existing legislative and academic efforts, through useful in providing diverse relevant insights to think through these issues, lack an overall, general framework accounting for these various aspects in a convincing manner.

In order to bridge this gap, we have introduced such a general framework, inspired by fundamental contributions in the philosophy of social sciences \cite{habermas1987com} and methodology of decision support \cite{meinard2019rationality}. The crux of our proposed general framework is a two-fold clarification: 
\begin{itemize}
\item a first clarification consists in distinguishing, on the one hand, \emph{explanations}, understood as accounts of the \emph{actual proceedings of the process} and, on the other hand, \emph{justifications}, defined as \emph{convincing and legitimate reasons why the process or its outcomes should be considered acceptable}.
\item a second clarification consists in distinguishing \emph{four models (technical, norm-oriented, expressive and communicative) of both explanations and justifications}, anchored in Habermas's typology of models of action.
\end{itemize}

A focal case study, the algorithmic decision process used to adjudicate admissions to French universities, was used to illustrate the usefulness and added-value of our proposed framework. Further attempts at exploring the promises and limits of our proposed framework are now needed to entrench its usefulness and/or refine its structure.

The article and works on the legal constraints of explainability of algorithmic decisions or their feasibility are complementary. Our focus here was not on legal regulation \emph{per se}. Our aim was rather to provide a framework of analysis that can shed light on legal norms concerning requirements articulated in terms of explainability and associated phrases.

\section{Acknowledgments}

The authors thank the Mission pour l'Interdisciplinarité et les Initatives Transervses of the CNRS (Paris) for financial support to the Fairness by Explanation of Algorithmic Decision (FEAD) and SPLEAD projects.

\bibliographystyle{splncs04}  
\bibliography{ref}

@book{perelman1958traite,
  title={Trait{\'e} de l'argumentation},
  author={Perelman, Cha{\"\i}m and Olbrechts-Tyteca, Lucie},
  volume={1},
  year={1958},
  publisher={Presses universitaires de France}
}

@book{Juanico-Sarles2020report,
  title={Assemblée nationale, Rapport d'information sur l'évaluation de l'accès à l'enseignement supérieur},
  author={Regis Juanico and Nathalie Sarles},
  volume={3232},
  year={2020},
  publisher={Assemblée nationale}
}

@book{Goltzberg2013argumentation,
  title={Cha{\"\i}m Perelman.L'argumentation juridique},
  author={Stefan Golltzberg},
  year={2013},
  publisher={Michalon - Le bien commun}
}

@book{boltanski2006justification,
  title={On justification: Economies of worth},
  author={Boltanski, Luc and Th{\'e}venot, Laurent},
  year={2006},
  publisher={Princeton University Press}
}

@book{habermas1987com,
    author = {Habermas, J{\"u}rgen},
    title = {The theory of communicative action},
    volume ={2: Lifeworld and system: A critique of functionalist reason},
    publisher = { Boston, MA: Bacon Press},
    year = {1987}
}

@article{miller2019explanation,
  title={Explanation in artificial intelligence: Insights from the social sciences},
  author={Miller, Tim},
  journal={Artificial intelligence},
  volume={267},
  pages={1--38},
  year={2019},
  publisher={Elsevier}
}

@article{walton2011explanation,
  title={A dialogue system specification for explanation},
  author={Walton, Douglas},
  journal={Synthese},
  volume={182},
  pages={349--374},
  year={2011},
  publisher={Springer}
}

@article{Hilton1990explanation,
  title={Conversational processes and causal explanation},
  author={Hilton, Denis J.},
  journal={Psychological Bulletin},
  volume={107},
  pages={65--81},
  year={1990},
  publisher={American Psychological Association}
}

@article{grice1975logic,
  title={Logic and conversation},
  author={Grice, Herbert Paul},
  journal={Syntax and semantics},
  volume={3},
  pages={43--58},
  year={1975}
}

@article{meinard2019rationality,
  title={On the rationality of decision aiding processes},
  author={Meinard, Yves and Tsouki{\`a}s, Alexis},
  journal={European Journal of Operational Research},
  volume={273},
  number={3},
  pages={1074--1084},
  year={2019},
  publisher={Elsevier}
}

@article{danaher2016threat,
  title={The threat of algocracy: Reality, resistance and accommodation},
  author={Danaher, John},
  journal={Philosophy \& technology},
  volume={29},
  number={3},
  pages={245--268},
  year={2016},
  publisher={Springer}
}

@article{waldman2019power,
  title={Power, process, and automated decision-making},
  author={Waldman, Ari Ezra},
  journal={Fordham L. Rev.},
  volume={88},
  pages={613},
  year={2019},
  publisher={HeinOnline}
}

@article{henin2022beyond,
	author = {Cl\'{e}ment Henin and Daniel Le M\'{e}tayer},
	doi = {10.1007/s00146-021-01251-8},
	journal = {AI and Society},
	number = {4},
	pages = {1397--1410},
	title = {Beyond Explainability: Justifiability and Contestability of Algorithmic Decision Systems},
	volume = {37},
	year = {2022}
}

@article{walton2007dialogical,
  title={Dialogical Models of Explanation.},
  author={Walton, Douglas},
  journal={ExaCt},
  volume={2007},
  pages={1--9},
  year={2007}
}

@article{henin2021framework,
  title={A framework to contest and justify algorithmic decisions},
  author={Henin, Cl{\'e}ment and Le M{\'e}tayer, Daniel},
  journal={AI and Ethics},
  volume={1},
  number={4},
  pages={463--476},
  year={2021},
  publisher={Springer}
}

@book{eubanks2018automating,
  title={Automating inequality: How high-tech tools profile, police, and punish the poor},
  author={Eubanks, Virginia},
  year={2018},
  publisher={St. Martin's Press}
}

@inproceedings{saxena2022unpacking,
  title={Unpacking invisible work practices, constraints, and latent power relationships in child welfare through casenote analysis},
  author={Saxena, Devansh and Moon, Seh Young and Shehata, Dahlia and Guha, Shion},
  booktitle={Proceedings of the 2022 CHI Conference on Human Factors in Computing Systems},
  pages={1--22},
  year={2022}
}

@article{stevenson2018assessing,
  title={Assessing risk assessment in action},
  author={Stevenson, Megan},
  journal={Minn. L. Rev.},
  volume={103},
  pages={303},
  year={2018},
  publisher={HeinOnline}
}

@inproceedings{haque2024are,
    author = {Haque, MD Romael and Saxena, Devansh and Weathington, Katy and Chudzik, Joseph and Guha, Shion},
    title = {Are We Asking the Right Questions?: Designing for Community Stakeholders’ Interactions with AI in Policing},
    year = {2024},
    isbn = {9798400703300},
    publisher = {Association for Computing Machinery},
    address = {New York, NY, USA},
    booktitle = {Proceedings of the 2024 CHI Conference on Human Factors in Computing Systems},
    articleno = {301},
    numpages = {20},
    location = {Honolulu, HI, USA},
    series = {CHI '24}
}

@article{burkart2021survey,
  title={A survey on the explainability of supervised machine learning},
  author={Burkart, Nadia and Huber, Marco F},
  journal={Journal of Artificial Intelligence Research},
  volume={70},
  pages={245--317},
  year={2021}
}

@article{gunning2019darpa,
  title={{DARPA’s} explainable artificial intelligence {(XAI)} program},
  author={Gunning, David and Aha, David},
  journal={AI magazine},
  volume={40},
  number={2},
  pages={44--58},
  year={2019}
}

@article{nauta2023anecdotal,
  title={From anecdotal evidence to quantitative evaluation methods: A systematic review on evaluating explainable ai},
  author={Nauta, Meike and Trienes, Jan and Pathak, Shreyasi and Nguyen, Elisa and Peters, Michelle and Schmitt, Yasmin and Schl{\"o}tterer, J{\"o}rg and Van Keulen, Maurice and Seifert, Christin},
  journal={ACM Computing Surveys},
  volume={55},
  number={13s},
  pages={1--42},
  year={2023},
  publisher={ACM New York, NY}
}

@article{arrieta2020explainable,
  title={Explainable Artificial Intelligence {(XAI)}: Concepts, taxonomies, opportunities and challenges toward responsible {AI}},
  author={Arrieta, Alejandro Barredo and D{\'\i}az-Rodr{\'\i}guez, Natalia and Del Ser, Javier and Bennetot, Adrien and Tabik, Siham and Barbado, Alberto and Garc{\'\i}a, Salvador and Gil-L{\'o}pez, Sergio and Molina, Daniel and Benjamins, Richard and others},
  journal={Information fusion},
  volume={58},
  pages={82--115},
  year={2020},
  publisher={Elsevier}
}

@article{guidotti2018survey,
  title={A survey of methods for explaining black box models},
  author={Guidotti, Riccardo and Monreale, Anna and Ruggieri, Salvatore and Turini, Franco and Giannotti, Fosca and Pedreschi, Dino},
  journal={ACM computing surveys (CSUR)},
  volume={51},
  number={5},
  pages={1--42},
  year={2018},
  publisher={ACM New York, NY, USA}
}

@article{adadi2018peeking,
  title={Peeking inside the black-box: a survey on explainable artificial intelligence {(XAI)}},
  author={Adadi, Amina and Berrada, Mohammed},
  journal={IEEE access},
  volume={6},
  pages={52138--52160},
  year={2018},
  publisher={IEEE}
}

@book{pearl2009causality,
  title={Causality},
  author={Pearl, Judea},
  year={2009},
  publisher={Cambridge university press}
}

@inproceedings{selvaraju2017grad,
  title={Grad-cam: Visual explanations from deep networks via gradient-based localization},
  author={Selvaraju, Ramprasaath R and Cogswell, Michael and Das, Abhishek and Vedantam, Ramakrishna and Parikh, Devi and Batra, Dhruv},
  booktitle={Proceedings of the IEEE international conference on computer vision},
  pages={618--626},
  year={2017}
}

@inproceedings{bhatt2020explainable,
  title={Explainable machine learning in deployment},
  author={Bhatt, Umang and Xiang, Alice and Sharma, Shubham and Weller, Adrian and Taly, Ankur and Jia, Yunhan and Ghosh, Joydeep and Puri, Ruchir and Moura, Jos{\'e} MF and Eckersley, Peter},
  booktitle={Proceedings of the 2020 conference on fairness, accountability, and transparency},
  pages={648--657},
  year={2020}
}

@inproceedings{kulesza2013too,
  title={Too much, too little, or just right? {Ways} explanations impact end users' mental models},
  author={Kulesza, Todd and Stumpf, Simone and Burnett, Margaret and Yang, Sherry and Kwan, Irwin and Wong, Weng-Keen},
  booktitle={2013 IEEE Symposium on visual languages and human centric computing},
  pages={3--10},
  year={2013},
  organization={IEEE}
}

@inproceedings{hohman2019gamut,
  title={Gamut: A design probe to understand how data scientists understand machine learning models},
  author={Hohman, Fred and Head, Andrew and Caruana, Rich and DeLine, Robert and Drucker, Steven M},
  booktitle={Proceedings of the 2019 CHI conference on human factors in computing systems},
  pages={1--13},
  year={2019}
}

@inproceedings{lakkaraju2016interpretable,
  title={Interpretable decision sets: A joint framework for description and prediction},
  author={Lakkaraju, Himabindu and Bach, Stephen H and Leskovec, Jure},
  booktitle={Proceedings of the 22nd ACM SIGKDD international conference on knowledge discovery and data mining},
  pages={1675--1684},
  year={2016}
}

@article{lundberg2017unified,
  title={A unified approach to interpreting model predictions},
  author={Lundberg, Scott M and Lee, Su-In},
  journal={Advances in neural information processing systems},
  volume={30},
  year={2017}
}

@article{bono2021algorithmic,
  title={Algorithmic fairness in credit scoring},
  author={Bono, Teresa and Croxson, Karen and Giles, Adam},
  journal={Oxford Review of Economic Policy},
  volume={37},
  number={3},
  pages={585--617},
  year={2021},
  publisher={Oxford University Press UK}
}

@article{obermeyer2019dissecting,
  title={Dissecting racial bias in an algorithm used to manage the health of populations},
  author={Obermeyer, Ziad and Powers, Brian and Vogeli, Christine and Mullainathan, Sendhil},
  journal={Science},
  volume={366},
  number={6464},
  pages={447--453},
  year={2019},
  publisher={American Association for the Advancement of Science}
}

@article{mohseni2021multidisciplinary,
  title={A multidisciplinary survey and framework for design and evaluation of explainable AI systems},
  author={Mohseni, Sina and Zarei, Niloofar and Ragan, Eric D},
  journal={ACM Transactions on Interactive Intelligent Systems (TiiS)},
  volume={11},
  number={3-4},
  pages={1--45},
  year={2021},
  publisher={ACM New York, NY}
}

@article{wick1992reconstructive,
  title={Reconstructive expert system explanation},
  author={Wick, Michael R and Thompson, William B},
  journal={Artificial Intelligence},
  volume={54},
  number={1-2},
  pages={33--70},
  year={1992},
  publisher={Elsevier}
}

@article{rudin2019stop,
  title={Stop explaining black box machine learning models for high stakes decisions and use interpretable models instead},
  author={Rudin, Cynthia},
  journal={Nature machine intelligence},
  volume={1},
  number={5},
  pages={206--215},
  year={2019},
  publisher={Nature Publishing Group UK London}
}

@article{green2019principles,
  title={The principles and limits of algorithm-in-the-loop decision making},
  author={Green, Ben and Chen, Yiling},
  journal={Proceedings of the ACM on Human-Computer Interaction},
  volume={3},
  number={CSCW},
  pages={1--24},
  year={2019},
  publisher={ACM New York, NY, USA}
}

@article{walton2006examination,
  title={Examination dialogue: An argumentation framework for critically questioning an expert opinion},
  author={Walton, Douglas},
  journal={Journal of pragmatics},
  volume={38},
  number={5},
  pages={745--777},
  year={2006},
  publisher={Elsevier}
}

@article{walton2011dialogue,
  title={A dialogue system specification for explanation},
  author={Walton, Douglas},
  journal={Synthese},
  volume={182},
  number={3},
  pages={349--374},
  year={2011},
  publisher={Springer}
}

@article{hilton1990conversational,
  title={Conversational processes and causal explanation.},
  author={Hilton, Denis J},
  journal={Psychological Bulletin},
  volume={107},
  number={1},
  pages={65},
  year={1990},
  publisher={American Psychological Association}
}

@article{caliskan2017semantics,
  title={Semantics derived automatically from language corpora contain human-like biases},
  author={Caliskan, Aylin and Bryson, Joanna J and Narayanan, Arvind},
  journal={Science},
  volume={356},
  number={6334},
  pages={183--186},
  year={2017},
  publisher={American Association for the Advancement of Science}
}

@article{mitchell2018prediction,
  title={Prediction-based decisions and fairness: A catalogue of choices, assumptions, and definitions},
  author={Mitchell, Shira and Potash, Eric and Barocas, Solon and D'Amour, Alexander and Lum, Kristian},
  journal={arXiv preprint arXiv:1811.07867},
  year={2018}
}

@article{hurlin2022fairness,
  title={The fairness of credit scoring models},
  author={Hurlin, Christophe and P{\'e}rignon, Christophe and Saurin, S{\'e}bastien},
  journal={arXiv preprint arXiv:2205.10200},
  year={2022}
}

@techreport{gunning2016explainable,
  title={Explainable artificial intelligence {(XAI)}}, 
  type = {darpa-baa-16-53},
  author={Gunning, David},
  institution={Defense Advanced Research Projects Agency},
  year={2016}
}

@article{mok2023people,
  title={People perceive algorithmic assessments as less fair and trustworthy than identical human assessments},
  author={Mok, Lillio and Nanda, Sasha and Anderson, Ashton},
  journal={Proceedings of the ACM on Human-Computer Interaction},
  volume={7},
  number={CSCW2},
  pages={1--26},
  year={2023},
  publisher={ACM New York, NY, USA}
}

@inproceedings{bates2024technology,
  title={Technology and Culture: How Predictive Policing Harmfully Profiles Marginalized People Groups},
  author={Bates, Taryn},
  booktitle={California Sociology Forum},
  volume={6},
  pages={18--27},
  year={2024}
}

@inproceedings{liu2018delayed,
  title={Delayed impact of fair machine learning},
  author={Liu, Lydia T and Dean, Sarah and Rolf, Esther and Simchowitz, Max and Hardt, Moritz},
  booktitle={International Conference on Machine Learning},
  pages={3150--3158},
  year={2018},
  organization={PMLR}
}

@article{wachter2017right,
  title={Why a right to explanation of automated decision-making does not exist in the general data protection regulation},
  author={Wachter, Sandra and Mittelstadt, Brent and Floridi, Luciano},
  journal={International data privacy law},
  volume={7},
  number={2},
  pages={76--99},
  year={2017},
  publisher={Oxford University Press}
}

@article{leake1995abduction,
  title={Abduction, experience, and goals: A model of everyday abductive explanation},
  author={Leake, David B},
  journal={Journal of Experimental \& Theoretical Artificial Intelligence},
  volume={7},
  number={4},
  pages={407--428},
  year={1995},
  publisher={Taylor \& Francis}
}

@inproceedings{ribeiro2016should,
  title={{"Why should i trust you?"} Explaining the predictions of any classifier},
  author={Ribeiro, Marco Tulio and Singh, Sameer and Guestrin, Carlos},
  booktitle={Proceedings of the 22nd ACM SIGKDD international conference on knowledge discovery and data mining},
  pages={1135--1144},
  year={2016}
}

@inproceedings{gilpin2018explaining,
  title={Explaining explanations: An overview of interpretability of machine learning},
  author={Gilpin, Leilani H and Bau, David and Yuan, Ben Z and Bajwa, Ayesha and Specter, Michael and Kagal, Lalana},
  booktitle={2018 IEEE 5th International Conference on data science and advanced analytics (DSAA)},
  pages={80--89},
  year={2018},
  organization={IEEE}
}

@article{doshi2017towards,
  title={Towards a rigorous science of interpretable machine learning},
  author={Doshi-Velez, Finale and Kim, Been},
  journal={arXiv preprint arXiv:1702.08608},
  year={2017}
}

@article{wachter2017counterfactual,
  title={Counterfactual explanations without opening the black box: Automated decisions and the {GDPR}},
  author={Wachter, Sandra and Mittelstadt, Brent and Russell, Chris},
  journal={Harv. JL \& Tech.},
  volume={31},
  pages={841},
  year={2017},
  publisher={HeinOnline}
}

@inproceedings{korikov2021counterfactual,
  title={Counterfactual explanations for optimization-based decisions in the context of the GDPR},
  author={Korikov, Anton and Shleyfman, Alexander and Beck, Chris},
  booktitle={ICAPS 2021 workshop on explainable AI planning},
  year={2021}
}

@article{harman1965inference,
  title={The inference to the best explanation},
  author={Harman, Gilbert H},
  journal={The philosophical review},
  volume={74},
  number={1},
  pages={88--95},
  year={1965},
  publisher={JSTOR}
}

@misc{aiact,
  author       = {{European Commission}},
  title        = {{EU AI Act}},
  howpublished = {Council of the EU, Press release},
  month        = {December},
  year         = {2023},
  note          = {https://artificialintelligenceact.eu/ai-act-explorer/}
}

@misc{gdpr,
  author       = {{European Parliament}},
  title        = {{General Data Protection Regulation}},
  howpublished = {Council of 27 April 2016 on the protection of natural persons with regard to the processing of personal data and on the free movement of such data},
  month        = {May},
  year         = {2016},
  note          = {http://data.europa.eu/eli/reg/2016/679/oj}
}

@article{perelman1961legal,
  title={Jugements de valeur, justification et argumentation},
  author={Perelman, Charles},
  journal={Revue Internationale de Philosophie},
  volume={15},
  pages={327--335},
  year={1961},
  publisher={Association Revue internationale de philosophie}
}

@article{rawls1997idea,
  title={The idea of public reason revisited},
  author={Rawls, John},
  journal={The university of Chicago law review},
  volume={64},
  number={3},
  pages={765--807},
  year={1997},
  publisher={JSTOR}
}

@misc{habermas1993justification,
  title={Justification and Application: Remarks on Discourse Ethics},
  author={Habermas, J{\"u}rgen},
  year={1993},
  publisher={MIT Press}
}

@article{binns2018algorithmic,
  title={Algorithmic accountability and public reason},
  author={Binns, Reuben},
  journal={Philosophy \& technology},
  volume={31},
  number={4},
  pages={543--556},
  year={2018},
  publisher={Springer}
}

@techreport{castelluccia2019understanding,
  title={Understanding algorithmic decision-making: Opportunities and challenges},
  author={Castelluccia, Claude and Le M{\'e}tayer, Daniel},
  year={2019},
  institution={European Parliament}
}

@article{fourcade2013classification,
  author  = {Marion Fourcade and Kieran Healy},
  title   = {Classification situations: Life-chances in the neoliberal era},
  journal = {Accounting, Organizations and Society},
  volume  = {38},
  number  = {8},
  pages   = {559--572},
  year    = {2013}
}

@inproceedings{kasy2021fairness,
  author    = {Maximilian Kasy and Rediet Abebe},
  title     = {Fairness, equality, and power in algorithmic decision-making},
  booktitle = {Proceedings of the 2021 ACM Conference on Fairness, Accountability, and Transparency (FAccT)},
  year      = {2021}
}

@book{gillespie2018custodians,
  author    = {Tarleton Gillespie},
  title     = {Custodians of the Internet: Platforms, Content Moderation, and the Hidden Decisions That Shape Social Media},
  publisher = {Yale University Press},
  year      = {2018}
}

@book{allouche2024contest,
  author    = {Annabelle Allouch and Delphine Espagnolo-Abadie},
  title     = {Contester Parcoursup},
  publisher = {Presses de Sciences Po},
  year      = {2024}
}

@article{edwards2018slave,
  author  = {Lilian Edwards and Michael Veale},
  title   = {Slave to the algorithm? Why a “right to an explanation” is probably not the remedy you are looking for},
  journal = {Duke Law \& Technology Review},
  volume  = {16},
  pages   = {18--84},
  year    = {2018}
}

@article{ananny2018seeing,
  author  = {Mike Ananny and Kate Crawford},
  title   = {Seeing without knowing: Limitations of the transparency ideal and its application to algorithmic accountability},
  journal = {New Media \& Society},
  volume  = {20},
  number  = {3},
  pages   = {973--989},
  year    = {2018}
}

@article{cinus2022effect, 
    title={The Effect of People Recommenders on Echo Chambers and Polarization}, 
    volume={16}, 
    number={1}, 
    journal={Proceedings of the International AAAI Conference on Web and Social Media}, 
    author={Cinus, Federico and Minici, Marco and Monti, Corrado and Bonchi, Francesco}, 
    year={2022}, 
    month={May}, 
    pages={90-101} 
}

@article{lipton2018mythos,
  title={The mythos of model interpretability: In machine learning, the concept of interpretability is both important and slippery.},
  author={Lipton, Zachary C},
  journal={Queue},
  volume={16},
  number={3},
  pages={31--57},
  year={2018},
  publisher={ACM New York, NY, USA}
}

@inproceedings{lanzetti2023impact,
  title={The impact of recommendation systems on opinion dynamics: Microscopic versus macroscopic effects},
  author={Lanzetti, Nicolas and D{\"o}rfler, Florian and Pagan, Nicol{\`o}},
  booktitle={2023 62nd IEEE Conference on Decision and Control (CDC)},
  pages={4824--4829},
  year={2023},
  organization={IEEE}
}

@article{parikh2019addressing,
  title={Addressing bias in artificial intelligence in health care},
  author={Parikh, Ravi B and Teeple, Stephanie and Navathe, Amol S},
  journal={Jama},
  volume={322},
  number={24},
  pages={2377--2378},
  year={2019},
  publisher={American Medical Association}
}

@article{Belghith2019orientation,
  title={L'orientation étudiante à l'heure de Parcours Sup. Des stratégies et des jugements socialement différenciés},
  author={Belghith Feres and Carvalho Huilton and Ferry Odile and Tenret \'Elise},
  journal={OVE Info},
   volume={39},
    year={2019}
}

@article{agarwal2023addressing,
  title={Addressing algorithmic bias and the perpetuation of health inequities: An {AI} bias aware framework},
  author={Agarwal, Ritu and Bjarnadottir, Margret and Rhue, Lauren and Dugas, Michelle and Crowley, Kenyon and Clark, Jessica and Gao, Gordon},
  journal={Health Policy and Technology},
  volume={12},
  number={1},
  pages={100702},
  year={2023},
  publisher={Elsevier}
}

@article{Gale1962algorithm,
title={College Admissions and the Stability of Marriage}, 
author={David Gale and Lloyd Shapley},
journal={The American Mathematical Monthly},
volume={69},
pages={9-15},
year={1962}
}

@article{ColorniTsoukias2024,
    author = {A. Colorni and A. Tsouki\`as},
    title = {What is a decision problem?},
    journal = {European Journal of Operational Research},
    volume = {314},
    pages = {255 - 267},
    year = {2024}
    }

@article{lerouge2026modeling,
  title={Modeling and generating user-centered contrastive explanations for the workforce scheduling and routing problem},
  author={Lerouge, Mathieu and Gicquel, C{\'e}line and Mousseau, Vincent and Ouerdane, Wassila},
  journal={International Transactions in Operational Research},
  volume={33},
  number={3},
  pages={1525--1558},
  year={2026},
  publisher={Wiley Online Library}
}

@book{ricoeur2014memoire,
  author    = {Ric{\oe}ur, Paul},
  title     = {La m{\'e}moire, l'histoire, l'oubli},
  publisher = {Le Seuil},
  address   = {Paris},
  series    = {L'Ordre philosophique},
  year      = {2014}
}

@book{vonwright1971,
    author={{G.H.} {von Wright}},
    title = {Explanation and Understanding},
    publisher = {Routledge},
    address = {London},
    year = {1971}
    }

@incollection{Lowe2009,
    author="Lowe, E. J.",
    editor="Sandis, Constantine",
    title="Free Agency, Causation and Action Explanation",
    bookTitle="New Essays on the Explanation of Action",
    year="2009",
    publisher="Palgrave Macmillan UK",
    address="London",
    pages="338--355",
    isbn="978-0-230-58297-2"
}
\end{document}